\documentclass[prd,reprint,amsmath,amssymb,superscriptaddress]{revtex4-2}
\newcommand{\be}{\begin{equation}}
\newcommand{\ee}{\end{equation}}
\newcommand{\ba}{\begin{align}}
\newcommand{\eda}{\end{align}}
\newcommand{\nn}{\nonumber}
\usepackage{bbm}
\usepackage{todonotes}\topmargin -1.2cm

\usepackage[export]{adjustbox}
\usepackage{graphicx}
\usepackage{epsfig}
\usepackage{braket}

\usepackage{bm}

\usepackage{capt-of}

\newcommand{\ppm}{\pi^{\pm}}

\newcommand{\hpm}{^{\pm}}

\newcommand{\tp}{\tilde{p}}
\newcommand{\uu}{_{1}}
\newcommand{\ud}{_{2}}
\newcommand{\ua}{_{a}}
\newcommand{\ub}{_{b}}
\newcommand{\us}{_{s}}
\newcommand{\2}{^{2}}
\newcommand{\cN}{\mathcal{N}}
\newcommand{\cM}{\mathcal{M}}

\newcommand{\ul}{_{\lambda}}

 \newcommand{\up}{_{p}}
  \newcommand{\ut}{_{t}}
 \newcommand{\upp}{_{\pi}}

  \def\er{\eqref}
  \newcommand{\bel}[1]{\be\label{#1}}
  \newcommand{\bal}[1]{\ba\label{#1}}
  \newcommand{\dv}{\text{d}}

  \renewcommand\slash[1]{\not \! #1}
  \newcommand{\tnp}{\slash{\tilde{p}}}
  
 \newcommand{\bp}{\slash{p}}
 \newcommand{\bk}{\slash{k}}

\usepackage{hyperref}

\numberwithin{equation}{section}
\renewcommand{\theequation}{\arabic{section}.\arabic{equation}}
\usepackage[noabbrev]{cleveref}    

\begin{document}

%\preprint{}

\title{Different versions of soft-photon theorems
exemplified at leading and next-to-leading terms for pion-pion and pion-proton scattering}
%Soft-photon theorem for pion-proton scattering: the next to leading term


\author{Piotr Lebiedowicz}
%\orcid{0000-0003-1963-6263}
\email{Piotr.Lebiedowicz@ifj.edu.pl}
\affiliation{Institute of Nuclear Physics Polish Academy of Sciences, 
Radzikowskiego 152, PL-31342 Krak{\'o}w, Poland}

\author{Otto Nachtmann}
\email{O.Nachtmann@thphys.uni-heidelberg.de}
\affiliation{Institut f\"ur Theoretische Physik, Universit\"at Heidelberg,
Philosophenweg 16, D-69120 Heidelberg, Germany}

\author{Antoni Szczurek}
%\orcid{0000-0001-5247-8442}
\email{Antoni.Szczurek@ifj.edu.pl}
\affiliation{Institute of Nuclear Physics Polish Academy of Sciences, 
Radzikowskiego 152, PL-31342 Krak{\'o}w, Poland}
\affiliation{College of Natural Sciences, 
Institute of Physics, University of Rzesz{\'o}w, 
Pigonia 1, PL-35310 Rzesz{\'o}w, Poland.}

%\date{\today}

\begin{abstract}
We investigate the photon emission in pion-pion and pion-proton scattering
in the soft-photon limit where the photon energy $\omega \to 0$.
The expansions of the $\pi^{-} \pi^{0} \to \pi^{-} \pi^{0} \gamma$ 
and the $\pi^{\pm} p \to \pi^{\pm} p \gamma$ amplitudes,
satisfying the energy-momentum relations,
to the orders $\omega^{-1}$ and $\omega^{0}$ are derived.
We show that these terms can be expressed completely
in terms of the on-shell amplitudes for 
$\pi^{-} \pi^{0} \to \pi^{-} \pi^{0}$ and $\pi^{\pm} p \to \pi^{\pm} p$,
respectively, and their partial derivatives with respect to \mbox{$s$ and $t$}.
The~structure term which is non singular for $\omega \to 0$ 
is determined to the order $\omega^{0}$ from the gauge-invariance constraint
using the generalized Ward identities for pions and the proton.

For the reaction $\pi^{-} \pi^{0} \to \pi^{-} \pi^{0} \gamma$ 
we discuss in detail the soft-photon theorems in the versions of both F.E. Low and S. Weinberg. We show that these two versions are different
and must not be confounded.
Weinberg's version gives the pole term of a Laurent expansion in $\omega$
of the amplitude for $\pi^{-} \pi^{0} \to \pi^{-} \pi^{0} \gamma$ around
the phase-space point of zero radiation.
Low's version gives an approximate expression for the above amplitude
at a fixed phase-space point, corresponding to non-zero radiation.
Clearly, the leading and next-to-leading terms 
in theses two approaches must be, and are indeed, different.
We show their relation.
We also discuss the expansions of differential cross sections
for $\pi^{-} \pi^{0} \to \pi^{-} \pi^{0} \gamma$ with respect to $\omega$
for $\omega \to 0$.

\end{abstract}

%\keywords{}

\maketitle

\section{Introduction}
\label{sec:1}
%--------------------------------------------------

In this article we shall discuss the production of soft photons in 
$\pi \pi$ and $\pi p$ scattering.
In particular, we shall study the following reactions
\begin{align}
\label{100}
\pi^{-} \,(p\ua)+ \pi^{0}\,(p\ub)&\to \pi^{-}\,(p\uu)+\pi^{0}\,(p\ud)\,, \\
\label{101}
\pi^{-} \,(p\ua)+ \pi^{0}\,(p\ub)&\to \pi^{-}\,(p'\uu)+\pi^{0}\,(p'\ud)
+\gamma\,(k,\varepsilon)\,,
\end{align}
and
\begin{align}
\label{1}
\ppm \,(p\ua)+ p\,(p\ub ,\lambda\ub )&\to \ppm\,(p\uu)+p\,(p\ud ,\lambda\ud )\,, \\
\label{2}
\ppm \,(p\ua)+ p\,(p\ub ,\lambda\ub )&\to \ppm\,(p'\uu)+p\,(p'\ud ,\lambda'\ud )+\gamma\,(k,\varepsilon)\,.
\end{align}
Here $p\ua, p\ub, p\uu, p\ud, p'\uu, p'\ud$ and $k$ are the momenta of the particles,
$\lambda\ub, \lambda\ud, \lambda'\ud$ are the spin indices of the protons, 
and $\varepsilon$ is the polarization vector of the photon.
Let $\omega=k^{0}$ be the photon energy in the overall c.m. system. We are interested in soft-photon production, 
$\omega\to 0$.

In a seminal paper F.E. Low \cite{Low:1958sn} derived the theorem 
that the leading term for $\omega\to 0$ 
in the soft-photon-production amplitudes 
comes from the emission of photons from the external particles of the reaction. 
In \cite{Low:1958sn} this was shown explicitly for the scattering of a charged scalar 
on an uncharged scalar particle,
that is, for a reaction like \er{101},
and for the scattering of a charged spin~1/2 fermion
on a neutral scalar boson.
In \cite{Weinberg:1964ew,Weinberg:1965nx} a soft-photon theorem was derived for general reactions
with an arbitrary number of external particles. 
In the following soft-photon production was studied by many authors;
see e.g. \cite{Gribov:1966hs,Burnett:1967km,Bell:1969yw,Liou:1978jx,Liou:1987ug,Lipatov:1988ii,
DelDuca:1990gz,
Korchin:1995ys,Korchin:1996up,
%Timmermans:2001cm,
Li:2011aq,
Gervais:2017yxv,Bern:2014vva,Lysov:2014csa,Bonocore:2021cbv,Engel:2021ccn,Engel:2023ifn}.

In \cite{Low:1958sn} also an expression for a next-to-leading term, 
of order $\omega^{0}$, is given for the scattering of scalars. In our study of soft-photon production in $\pi^{-}\pi^{0}$ scattering we recalculated the next to leading term and found a different result \cite{Lebiedowicz:2021byo}. 
In the present paper, 
we reconsider the leading and next-to-leading terms 
in the soft-photon expansion of \er{101}.
We shall show that Low's version \cite{Low:1958sn} and Weinberg's version \cite{Weinberg:1965nx}
of the soft-photon theorem are different and should not be confused.
We also shall show how they are related.
In this way we shall also give the clear reason why there must be
a difference between the formulas for the next-to-leading terms
in Low's paper \cite{Low:1958sn} and in our paper \cite{Lebiedowicz:2021byo}
and how the two results are related.
These results have already been presented in a short form at a conference in October 2023;
see \cite{Nachtmann_talk}.
There, also our misunderstanding of the results of \cite{Low:1958sn}
which led us to an error in \cite{Lebiedowicz:2021byo}
was corrected. 
This will be discussed in detail below.

Our paper is organized as follows.
In Sec.~\ref{sec:2} we discuss the phase space 
and the kinematics for the reactions \er{100} and \er{101}.
Section~\ref{sec:3} deals with the reactions \er{100} and \er{101}
from a general point of view.
In Sec.~\ref{sec:4} we recall the expansion of the amplitude
for $\pi^{-} \pi^{0} \to \pi^{-} \pi^{0} \gamma$
around the phase-space point $p_{1}, p_{2}, k = 0$
as presented in \cite{Lebiedowicz:2021byo}, 
where the leading term is precisely given by the soft-photon theorem
of \cite{Weinberg:1965nx}.
Section~\ref{sec:5} deals with Low's version of the soft-photon theorem 
\cite{Low:1958sn}
and its relation to the results of \cite{Weinberg:1965nx,Lebiedowicz:2021byo}.
In Sec.~\ref{sec:6} we discuss cross sections for
$\pi^{-} \pi^{0} \to \pi^{-} \pi^{0} \gamma$.
In Sec.~\ref{sec:7} we give an outline of the calculation for
the leading and next-to-leading terms of the 
$\pi^{\pm} p \to \pi^{\pm} p \gamma$ reactions \er{2}.
Section~\ref{sec:8} contains a summary and our conclusions.
Some details of our analysis are given 
in Appendices~\ref{sec:appendixA} and \ref{sec:appendixB}.

In our paper we use the following theoretical framework
for the calculations.
We consider the reactions \er{100}--\er{2} in QCD plus
leading order in electromagnetism.
We use only exact Quantum Field Theory (QFT) methods in this framework:
\begin{itemize}
\item energy-momentum conservation,
\item gauge invariance,
\item parity ($P$), charge conjugation ($C$), and time-reversal ($T$) invariance,\item the generalized Ward identity for the pion
and the proton fields,
which in QCD are composite local fields,
\item analyticity properties of amplitudes,
in particular the Landau conditions.
\end{itemize}

It turned out that the evaluation of the $\omega^{0}$ term 
for the amplitude of~\er{2} involved a lengthy and complex analysis. 
Therefore, we present in this paper only the basic ingredients of the calculation and the results. 
All details can be found in the companion paper \cite{Lebiedowicz:2023mlz}.

%--------------------------------------------------
\section{Kinematics and phase space for $\pi \pi \to \pi \pi$ and $\pi \pi \to \pi \pi \gamma$}
\label{sec:2}
%--------------------------------------------------

\begin{figure}[!ht]
\includegraphics[width=.38\textwidth]{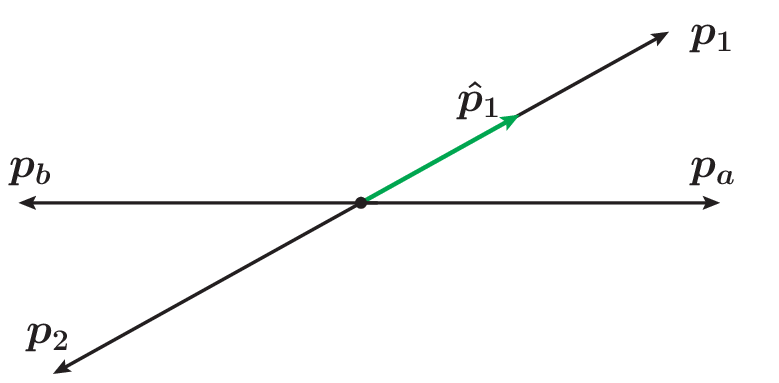}
\caption{The reaction 
$\pi^{-} \pi^{0} \to \pi^{-} \pi^{0}$ (\ref{2.1})
in the c.m. system.}
\label{fig:1}
\end{figure}

Let us start with the elastic reaction
\begin{eqnarray}
&&\pi^{-} \,(p_{a}) + \pi^{0} \,(p_{b}) \to \pi^{-} \,(p_{1}) + \pi^{0} \,(p_{2})\,, \nonumber\\
&&p_{a} + p_{b} = p_{1} + p_{2}\,.
\label{2.1}
\end{eqnarray}
We set as usual
\begin{eqnarray}
&& s = (p_{a} + p_{b})^{2} = (p_{1} + p_{2})^{2}\,, \nonumber \\
&& t = (p_{a} - p_{1})^{2} = (p_{b} - p_{2})^{2}\,.
\label{2.2}
\end{eqnarray}
Let us look at the reaction (\ref{2.1}) in the c.m. system
and consider a given value of the c.m. energy squared $s$.
Then the energies and absolute values of the momenta are fixed,
\begin{eqnarray}
&& p_{a}^{0} = p_{b}^{0} = p_{1}^{0} = p_{2}^{0} = \frac{\sqrt{s}}{2}\,, \nonumber\\
&& |\bm{p_{a}}| = |\bm{p_{b}}| = |\bm{p_{1}}| = |\bm{p_{2}}|
=  \sqrt{\frac{s}{4} - m_{\pi}^{2}}\,.
\label{2.3}
\end{eqnarray}
For a given initial configuration we can vary only
\mbox{$\bm{\hat{p}\uu}=\bm{p\uu}/|\bm{p\uu}|$},
the unit vector in direction of $\bm{p\uu}$;
see Fig.~\ref{fig:1}.
The phase space is the unit sphere.

\begin{figure}[!ht]
\includegraphics[width=.38\textwidth]{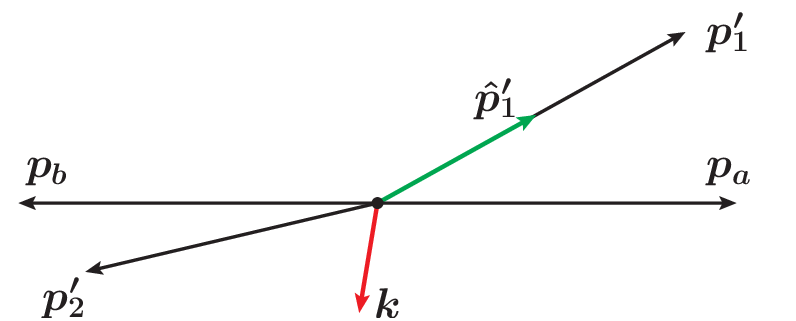}
\caption{The reaction 
$\pi^{-} \pi^{0} \to \pi^{-} \pi^{0} \gamma$ (\ref{2.4})
in the c.m. system.}
\label{fig:2}
\end{figure}

Now we go to the reaction with photon radiation
\begin{eqnarray}
&&\pi^{-} \,(p_{a}) + \pi^{0} \,(p_{b}) \to \pi^{-} \,(p_{1}') + \pi^{0} \,(p_{2}')
+ \gamma \,(k,\varepsilon)\,, \nonumber\\
&&p_{a} + p_{b} = p'_{1} + p'_{2} + k\,.
\label{2.4}
\end{eqnarray}
Here we set
\begin{eqnarray}
&& s = (p_{a} + p_{b})^{2} = (p_{1}' + p_{2}' + k)^{2}\,, \nonumber \\
&& t_{1} = (p_{a} - p_{1}')^{2} = (p_{b} - p_{2}' - k)^{2}\,, \nonumber \\
&& t_{2} = (p_{b} - p_{2}')^{2} = (p_{a} - p_{1}' - k)^{2}\,.
\label{2.5}
\end{eqnarray}
We shall consider real and virtual photon emission and require
\begin{eqnarray}
k^{2} \geqslant 0\,, \quad k^{0} \geqslant 0\,,
\label{2.6}
\end{eqnarray}
and $k$ small, say $|k^{\mu}| \ll \sqrt{s - 4 m_{\pi}^{2}}$
$(\mu = 0, \ldots, 3)$
in the c.m. system.
We consider a given value of $s$ and ask what are the free parameters
of the reaction (\ref{2.4}).
In the c.m. system (Fig.~\ref{fig:2}) 
a convenient set of such parameters
is given by the four-vector $k$ plus the unit vector
$\bm{\hat{p}\uu'}=\bm{p\uu'}/|\bm{p\uu'}|$:
\begin{align}
{\rm Phase \;space \;of \;(\ref{2.4})} 
= \lbrace \left( k, \bm{\hat{p}\uu'} \right), \,
k \in {\rm part \;of} \,{\rm R}_{4}, \,|\bm{\hat{p}\uu'}| = 1 \rbrace \,.
\label{2.7}
\end{align}

We can easily see this by considering
the reaction (\ref{2.4}) for given $k$
in the rest system of the four-vector 
\mbox{$p_{a} + p_{b} - k$}.
In this system $\bm{p\uu'}$ and $\bm{p\ud'}$
are back-to-back with fixed $|\bm{p\uu'}|$ 
and $|\bm{p\ud'}|$.
The only freedom left is to vary $\bm{p\uu'}$
in any direction. 
The same is then also true in the c.m. system if $k$
is small enough.

%--------------------------------------------------
\section{General analysis of $\pi \pi \to \pi \pi$ and $\pi \pi \to \pi \pi \gamma$}
\label{sec:3}
%--------------------------------------------------

\begin{figure}[!h]
\includegraphics[width=.23\textwidth]{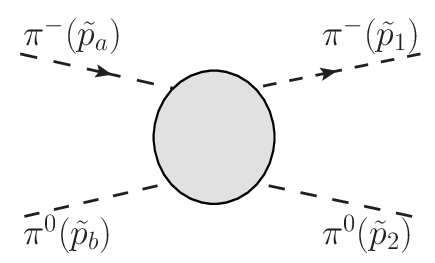}
\caption{Diagram for the off-shell and on-shell reaction (\ref{3.1}).}
\label{fig:3}
\end{figure}

We consider first the reaction
\begin{equation}
\pi^{-}\,(\tp\ua)+ \pi^{0} \,(\tp\ub)\to \pi^{-}\,(\tp\uu)+\pi^{0} \,(\tp\ud)
\label{3.1}
\end{equation}
off shell and on shell.
We have always energy-momentum conservation
\begin{equation}
\tp\ua +\tp\ub =\tp\uu +\tp \ud\,.
\label{3.2}
\end{equation}
In relations which hold off shell and on shell we denote momenta with a tilde.
The diagram for (\ref{3.1}) is shown in Fig.~\ref{fig:3}.
%-------------------------------------------------------------
%-------------------------------------------------------------

As kinematic variables we have the masses of the, 
in general off-shell, pions, an energy variable, and a $t$-variable:
\begin{eqnarray}
&&\tilde{\nu} = \tp_{a} \cdot \tp_{b} + \tp_{1} \cdot \tp_{2}\,,
\nonumber \\
&&\tilde{t} = (\tp_{a} - \tp_{1})^{2} = (\tp_{b} - \tp_{2})^{2}\,,
\nonumber \\
&&m_{a}^{2}=\tp_{a}^{2}\,,\quad
m_{b}^{2}=\tp_{b}^{2}\,,\quad
m_{1}^{2}=\tp_{1}^{2}\,,\quad
m_{2}^{2}=\tp_{2}^{2}\,.\qquad
\label{3.3}
\end{eqnarray}
We use here, following Low, $\tilde{\nu}$ 
as energy variable.
For the Mandelstam $\tilde{s}$ variable we get
\begin{eqnarray}
\tilde{s} = (\tp_{a} + \tp_{b})^{2} = 
\tilde{\nu} + \frac{1}{2} \left(m_{a}^{2} + m_{b}^{2} + m_{1}^{2} + m_{2}^{2} \right)\,. \qquad
\label{3.4}
\end{eqnarray}
The scattering amplitude for (\ref{3.1})
can only depend on the variables (\ref{3.3}):
\begin{eqnarray}
{\cal T}(\tp_{1},\tp_{2},\tp_{a},\tp_{b}) =
{\cal M}(\tilde{\nu},\tilde{t},m_{a}^{2},m_{b}^{2},m_{1}^{2},m_{2}^{2})\,.
\label{3.5}
\end{eqnarray}
The on-shell amplitude is obtained setting
\begin{eqnarray}
&&\tp_{a} \to p_{a}\,,
\tp_{b} \to p_{b}\,,
\tp_{1} \to p_{1}\,,
\tp_{2} \to p_{2}\,,
\nonumber \\
&&m_{a}^{2}=m_{b}^{2}=
m_{1}^{2}=m_{2}^{2}=m_{\pi}^{2}\,,
\nonumber \\
&& \tilde{\nu} \to \nu, \quad \tilde{t} \to t\,,
\label{3.6}
\end{eqnarray}
and we get
\begin{eqnarray}
\left.{\cal T}(p_{1},p_{2},p_{a},p_{b})\right|_{\rm on\; shell} &=&
{\cal M}(\nu, t, m_{\pi}^{2},m_{\pi}^{2},m_{\pi}^{2},m_{\pi}^{2})
\nonumber \\
&\equiv& {\cal M}^{(\rm on)}(\nu, t)\,.
\label{3.6a}
\end{eqnarray}

Next we come to the photon-emission reaction (\ref{101}):
\begin{align}
\pi^{-} \,(p\ua)+ \pi^{0}\,(p\ub)&\to \pi^{-}\,(p'\uu)+\pi^{0}\,(p'\ud)
+\gamma\,(k,\varepsilon)\,,
\label{3.7}
\end{align}
where energy-momentum conservation reads 
\begin{eqnarray}
p_{a} + p_{b} = p'_{1} + p'_{2} + k\,.
\label{3.8}
\end{eqnarray}
Note that for four-vector $k \neq 0$ we \textit{must have}
$p'_{1} \neq p_{1}$, $p'_{2} \neq p_{2}$
with $p_{1}$, $p_{2}$ from (\ref{2.1}).
The amplitude for (\ref{3.7}) is
\bal{3.9}
&\braket{\pi^{-} (p'\uu), \, \pi^{0}(p'\ud),\, \gamma (k,\varepsilon)|{\mathcal T}|\pi^{-} (p\ua),\, \pi^{0}(p\ub)}\nn\\
&\quad =(\varepsilon^{\lambda})^{*} 
{\cal M}_{\lambda}(p'\uu , p'\ud , k , p\ua , p\ub) \,.
\end{align}
In the following we consider ${\cal M}_{\lambda}$
for real and also for virtual, timelike, photons, that is,
for (\ref{2.6}).

\begin{figure}[!h]
(a)\includegraphics[width=.3\textwidth]{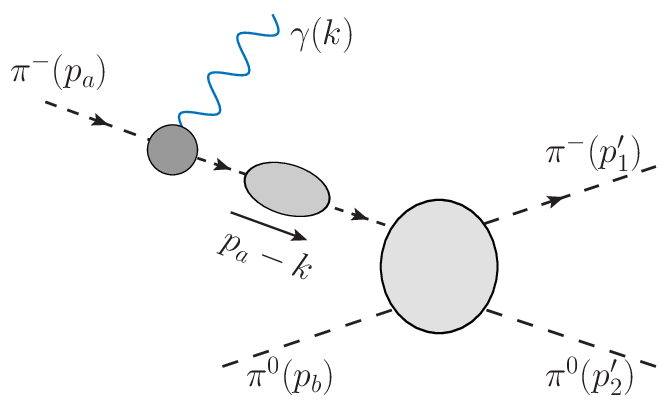}\\
(b)\includegraphics[width=.3\textwidth]{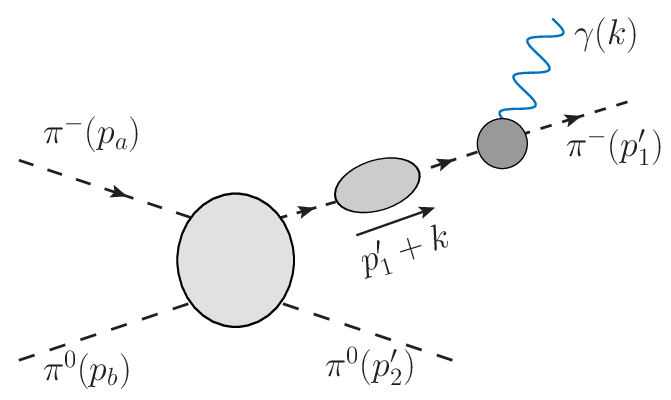}\\
(c)\qquad \includegraphics[width=.26\textwidth]{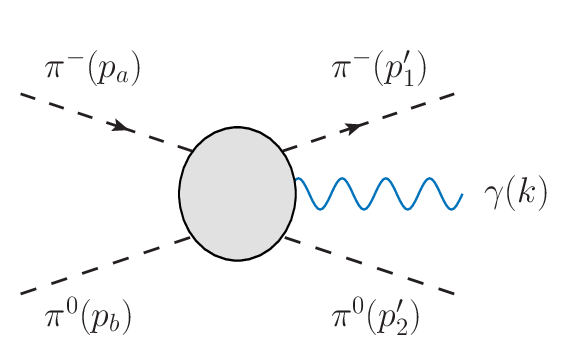}
\caption{Diagrams for the reaction 
$\pi^{-} \pi^{0} \to \pi^{-} \pi^{0} \gamma$ (\ref{3.7}).
The diagrams (a) and (b) describe the photon emission from the external charged lines,
the diagram (c) corresponds to the photon emission from internal lines,
the structure term.
The blobs in (a) and (b) stand for the full pion propagator,
the full $\gamma \pi \pi$ vertex function, and the off-shell
$\pi \pi$ scattering amplitude.}
\label{fig:4}
\end{figure}

There are three diagrams for the reaction (\ref{3.7})
as shown in Fig.~\ref{fig:4},
two one-particle reducible ones, (a) and (b),
and one which is one-particle irreducible (c).
%-------------------------------------------------------------
%-------------------------------------------------------------

To calculate the diagrams (a) and (b) we need the off-shell
$\pi \pi \to \pi \pi$ amplitude which we have already introduced,
the pion propagator $\Delta_{F}(p^{2})$, and the pion-photon vertex function
$\widehat{\Gamma}\ul^{(\gamma \pi\pi)}(p',p)$:
\begin{eqnarray}
\includegraphics[width=.22\textwidth]{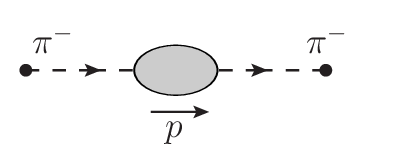}
&&i\Delta_{F}(p^{2})\,,
\label{3.10}\\
\includegraphics[width=.22\textwidth]{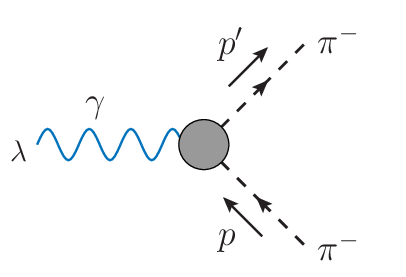}
&&ie\widehat{\Gamma}\ul^{(\gamma \pi\pi)}(p',p)\,. \quad
\label{3.11}
\end{eqnarray}
We denote by $e = \sqrt{4 \pi \alpha_{\rm em}} > 0$ the $\pi^{+}$ charge.

We get
\begin{eqnarray}
{\cal M}_{\lambda} =
{\cal M}_{\lambda}^{(a)} + 
{\cal M}_{\lambda}^{(b)} + 
{\cal M}_{\lambda}^{(c)}\,,
\label{3.11a}
\end{eqnarray}
where
\begin{eqnarray}
{\cal M}_{\lambda}^{(a)} &=& -e \,{\cal M}^{(a)}\, \Delta_{F}[ (p_{a}-k)^{2} ]\,
\widehat{\Gamma}\ul^{(\gamma \pi\pi)}(p_{a}-k,p_{a})\,, \nonumber \\
{\cal M}^{(a)} &=& 
\left.{\cal T}(p_{1}',p_{2}',p_{a}-k,p_{b})\right|_{\rm off\; shell}  \nonumber \\
&=&
{\cal M}[ (p_{a}-k, p_{b}) + p_{1}' \cdot p_{2}',  \nonumber \\
&& \qquad (p_{b}-p_{2}')^{2}, (p_{a}-k)^{2}, m_{\pi}^{2}, m_{\pi}^{2}, m_{\pi}^{2} ]\,, \nonumber \\
\label{3.12}
\end{eqnarray}
\begin{eqnarray}
{\cal M}_{\lambda}^{(b)} &=& -e \,
\widehat{\Gamma}\ul^{(\gamma \pi\pi)}(p_{1}',p_{1}'+k)\,
\Delta_{F}[ (p_{1}'+k)^{2} ]\, {\cal M}^{(b)}\,,
\nonumber \\
{\cal M}^{(b)} &=& 
\left.{\cal T}(p_{1}'+k,p_{2}',p_{a},p_{b})\right|_{\rm off\; shell}  \nonumber \\
&=&
{\cal M}[ (p_{a} \cdot p_{b}) + (p_{1}' + k, p_{2}'),  \nonumber \\
&& \qquad (p_{b}-p_{2}')^{2}, m_{\pi}^{2}, m_{\pi}^{2}, (p_{1}'+k)^{2}, m_{\pi}^{2} ]\,. \nonumber \\
\label{3.13}
\end{eqnarray}

We shall now use one of the best tools from QFT which
we have:
\textit{gauge invariance}.
This gives us the generalized Ward identity \cite{Ward:1950xp,Takahashi:1957xn}
\begin{eqnarray}
(p'-p)^{\lambda} \widehat{\Gamma}\ul^{(\gamma \pi\pi)}(p',p) =
\Delta_{F}^{-1}(p'^{2}) - \Delta_{F}^{-1}(p^{2}) \quad
\label{3.14}
\end{eqnarray}
and the condition
\begin{eqnarray}
k^{\lambda}{\cal M}_{\lambda} =
k^{\lambda} \left( {\cal M}_{\lambda}^{(a)} + 
{\cal M}_{\lambda}^{(b)} + 
{\cal M}_{\lambda}^{(c)} \right) = 0 \,.
\label{3.15}
\end{eqnarray}
From these two conditions we get an exact relation
between
${\cal M}_{\lambda}^{(c)}$, and ${\cal M}^{(a)}$, 
${\cal M}^{(b)}$:
\begin{eqnarray}
k^{\lambda} {\cal M}_{\lambda}^{(c)} = -e\,{\cal M}^{(a)}
                                       +e\,{\cal M}^{(b)}\,;
\label{3.16}
\end{eqnarray}
cf.~(2.22) of \cite{Lebiedowicz:2021byo}.
These are the QFT relations which we shall use in the following.

%--------------------------------------------------
\section{Soft photon theorem I}
\label{sec:4}
%--------------------------------------------------

\begin{figure}[!h]
\includegraphics[width=.25\textwidth]{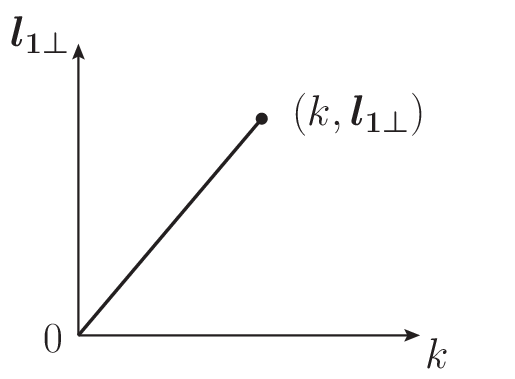}
\caption{Schematic representation of the six-dimensional
neighbourhood of the phase-space-point
$(k = 0, \bm{\hat{p}\uu'} = \bm{\hat{p}\uu})$;
see (\ref{4.1}).}
\label{fig:5}
\end{figure}

In this section we shall give the expansion of the amplitude
${\cal M}_{\lambda}$ around the phase-space point
$(k = 0, \bm{\hat{p}\uu'} = \bm{\hat{p}\uu})$.
In a small neighbourhood of this phase-space point
we set, assuming $|\bm{l_{1 \perp}}| = {\mathcal O}(\omega)$,
\begin{eqnarray}
\bm{\hat{p}\uu'} = \bm{\hat{p}\uu} - 
\frac{\bm{l_{1 \perp}}}{|\bm{p_{1}}|}\,,
\quad 
\bm{l_{1 \perp}} \cdot \bm{\hat{p}\uu} = 
0 + {\mathcal O}(\omega^{2})\,.
\label{4.1}
\end{eqnarray}
This neighbourhood has 6 dimensions, schematically we represent it
as shown in Fig.~\ref{fig:5}.
%-------------------------------------------------------------
%-------------------------------------------------------------

For $(k = 0, \bm{l_{1 \perp}} = 0)$ we have the kinematics of
$\pi^{-} \pi^{0} \to \pi^{-} \pi^{0}$,
the reaction without radiation (\ref{2.1}).
For $(k, \bm{l_{1 \perp}})$ we have
\begin{equation}
p'\uu=p\uu-l\uu\,,\quad p'\ud=p\ud-l\ud\,,
\label{4.2}
\end{equation}
where in the c.m. frame we get after
a simple calculation, up to order $\omega$,
the following result.
%Energy-momentum conservation reads here
%\bel{11}
%p\ua +p\ub = p'\uu + p'\ud + k\,.
%\ee
%Clearly, for $k \neq 0$ we must have $(p'\uu , p'\ud )\neq (p\uu , p\ud)$; see~\er{3}. 
%Therefore, we write 
%\bel{12}
%p'\uu=p\uu-l\uu\,,\quad p'\ud=p\ud-l\ud\,,
%\ee
%and determine $l\uu$, $l\ud$ from
%\begin{align}\label{13}
%l\uu+l\ud&=k\,,\nn\\
%(p\uu-l\uu)\2&=m_{\pi}\2\,,\nn\\
%(p\ud-l\ud)\2&=m_{\pi}\2\,.
%\end{align}
%We are interested in $k\to 0$ and assume, therefore, %also small $l_{1,2}$.
%Then \er{13} is easily solved for $l_{1,2}$.
We have in the overall c.m. system 
of (\ref{2.1}) and (\ref{2.4}) with 
$\bm{\hat{p}\uu}=\bm{p\uu}/|\bm{p\uu}|$,
\begin{align}
\label{14}
&(p_{1}^{\mu})=
\left(\begin{array}{l}
p_{1}^{0}
\vspace{0.1cm}\\
|\bm{p_{1}}|\,\bm{\hat{p}\uu}
\end{array}\right), \quad
(p_{2}^{\mu})=
\left(\begin{array}{l}
p_{2}^{0}
\vspace{0.1cm}\\
-|\bm{p_{2}}|\,\bm{\hat{p}\uu}
\end{array}\right), \nn\\
&(k^{\mu})=
\left( \begin{array}{l}
k^{0}
\vspace{0.1cm}\\
k_{\parallel}\,\bm{\hat{p}\uu}+\bm{k_{\perp}}
\end{array}\right), 
\quad
\bm{k_{\perp}} \cdot \bm{\hat{p}\uu}=0\,,
\end{align}
where $k^{0}=\omega$, and $p_{1,2}^{0}$, $|\bm{p_{1,2}}|$
are as in (\ref{2.3}).
We find then [see (3.21) of \cite{Lebiedowicz:2021byo}]
\begin{align}
\label{15}
(l\uu^{\mu})&=\left( \begin{array}{l}
\dfrac{1}{\sqrt{s}} (p\ud\cdot k)
\vspace{0.1cm}\\
\dfrac{p_{1}^{0}}{|\bm{p\uu}|\sqrt{s}} (p\ud\cdot k) \bm{\hat{p}\uu} +\bm{l_{1 \perp}}
\end{array}\right),
\nn\\
(l\ud^{\mu})&=\left( \begin{array}{l}
\dfrac{1}{\sqrt{s}} (p\uu\cdot k)
\vspace{0.1cm}\\
\bm{k}-\dfrac{p_{1}^{0}}{|\bm{p\uu}|\sqrt{s}} (p\ud\cdot k) \bm{\hat{p}\uu} -\bm{l_{1 \perp}}
\end{array}\right).
%\nn\\
%\bm{l_{1 \perp}} \cdot \hat{p}_{1} &=0\,.
\end{align}
We have written (\ref{15}) in such a way
that it holds, of course, for $\pi \pi$ scattering,
inserting for $s$, $p_{1,2}$, $|\bm{p\uu}|$,
and $\bm{\hat{p}\uu} = \bm{p\uu}/|\bm{p\uu}|$
the expressions from (\ref{2.2}), (\ref{2.3}),
and (\ref{14}).
We shall see in Sec.~\ref{sec:7} that for
the $\pi p$ scattering case the analogous quantities
$l_{1,2}$ are again given by (\ref{15})
with the appropriate expressions for 
$s$, $p_{1,2}$, $|\bm{p\uu}|$,
and $\bm{\hat{p}\uu}$ inserted; 
see Eq.~(\ref{7.100}) ff.

\begin{figure}[!ht]
\includegraphics[width=.35\textwidth]{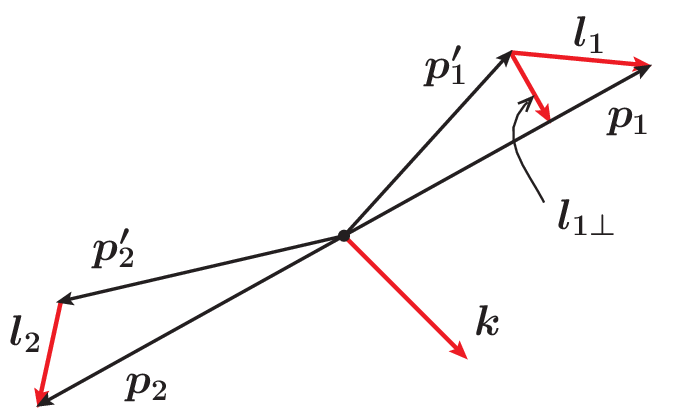}
\caption{
The final states of (\ref{2.1}) and (\ref{2.4}) 
for $|\bm{l_{1 \perp}}| = {\mathcal O}(\omega)$
in the c.m. system.}
\label{fig:6}
\end{figure}

We illustrate the situation in the overall c.m. system
in Fig.~\ref{fig:6}.
%--------------------------------------------
%--------------------------------------------

We have the important relations (up to order $\omega$)
\begin{align}
\label{4.4}
l\uu+l\ud &= k\,,\nn\\
p\uu\cdot l\uu &= p\ud \cdot l\ud=0\,.
\end{align}

Now we come to the expansion of the amplitude
${\cal M}_{\lambda}$ for $\omega \to 0$.
To be precise we set, in the c.m. system,
\begin{align}
\label{4.5}
(k^{\mu})=
\omega \left( \begin{array}{l}
1
\vspace{0.1cm}\\
\bm{\tilde{k}}
\end{array}\right), \quad 
\omega \geqslant 0, \quad
\bm{\tilde{k}}^{2} \leqslant 1\,.
\end{align}
In this way we have always 
\begin{align}
\label{4.6a}
k^{2}=
\omega^{2}(1 - \bm{\tilde{k}}^{2}) \geqslant 0\,.
\end{align}
Furthermore we set
\begin{align}
\label{4.6}
\bm{l_{1 \perp}} = \omega \;\bm{\tilde{l}_{1 \perp}}\,,
\quad |\bm{\tilde{l}_{1 \perp}}| = {\mathcal O}(1)\,.
\end{align}
We keep $\bm{\tilde{k}}$ and $\bm{\tilde{l}_{1 \perp}}$ fixed
and consider the expansion of the radiative amplitude
for $\omega \to 0$.
That is, we consider ${\cal M}_{\lambda}$ on a line
starting from the origin
in the phase space shown schematically in Fig.~\ref{fig:5}.
Of course, this will be a \textit{Laurent expansion}.

Note that when constructing this expansion we have to count
$k^{2} = {\mathcal O}(\omega^{2})$ due to (\ref{4.6a}).
In theoretical considerations we certainly can consider $k^{\mu}$ (\ref{4.5})
and $k^{2}$ (\ref{4.6a}) for fixed $\bm{\tilde{k}}$ 
and $\omega \to 0$. If we want to realize $k^{2} > 0$ in nature
by virtual photon $\gamma^{*}$-production with 
$\gamma^{*} \to e^{+} e^{-}$ we have, of course, the limit
\begin{align}
\label{4.7a}
\omega&\geqslant 2 m_{e}\,, \nn \\
k^{2}&=
\omega^{2}(1 - \bm{\tilde{k}}^{2}) \geqslant 4 m_{e}^{2}\,,
\end{align}
where $m_{e}$ is the electron mass. But the electron mass
is very small on a hadronic scale. 
Thus, in $\gamma^{*}$-production with the decay 
$\gamma^{*} \to e^{+} e^{-}$ we can reach very low values
of $\omega$ and $k^{2}$ where the soft-photon expansion has
a good chance to be valid.

Now we illustrate the construction of the Laurent expansion
for the term ${\cal M}_{\lambda}^{(a)}$;
see Fig.~\ref{fig:4}(a) and (\ref{3.12}).
We have
\begin{eqnarray}
{\cal M}_{\lambda}^{(a)} &=& -e
\left.{\cal T}(p_{1}',p_{2}',p_{a}-k,p_{b})\right|_{\rm off\; shell}
\nonumber \\
&& \times
\Delta_{F}[ (p_{a}-k)^{2} ]\,
\widehat{\Gamma}\ul^{(\gamma \pi\pi)}(p_{a}-k,p_{a})\,.
\label{4.7}
\end{eqnarray}
Note that our off-shell amplitude satisfies
energy-momentum conservation:
\begin{eqnarray}
p_{a} - k + p_{b} = p'_{1} + p'_{2}\,.
\label{4.8}
\end{eqnarray}
From the generalized Ward identity (\ref{3.14})
we find for $\omega \to 0$ 
[see (3.13) of \cite{Lebiedowicz:2021byo}]:
\bal{4.9}
&\Delta_{F}[ (p\ua-k)^{2} ]\, 
\widehat{\Gamma}\ul^{(\gamma \pi\pi)}(p\ua-k,\, p\ua)\nn\\
&\quad =\frac{(2p\ua -k)\ul}{-2p\ua\cdot k + k\2 +i\varepsilon}+{\mathcal O}(\omega)\,.
\end{align}
This shows that in order to get the expansion for
${\cal M}_{\lambda}^{(a)}$ to the orders
$\omega^{-1}$ and $\omega^{0}$
we have to calculate the expansion of
\begin{eqnarray}
&&\left.{\cal T}(p_{1}',p_{2}',p_{a}-k,p_{b})\right|_{\rm off\; shell} \nonumber \\
&&\quad =
\left.{\cal T}(p_{1}-l_{1},p_{2}-l_{2},p_{a}-k,p_{b})\right|_{\rm off\; shell} \nonumber\\
&&\quad ={\cal M}[ (p_{a}-k, p_{b}) + (p_{1} - l_{1}, 
p_{2} - l_{2}),  \nonumber \\
&& \qquad \quad (p_{b}-p_{2}+l_{2})^{2}, (p_{a}-k)^{2}, m_{\pi}^{2}, m_{\pi}^{2}, m_{\pi}^{2} ] \qquad
\label{4.10}
\end{eqnarray}
to the orders $\omega^{0}$ and $\omega$.
Note that, of course, $l_{1}$ and $l_{2}$ have to be taken into account.
Treating in this way ${\cal M}_{\lambda}^{(a)}$
and ${\cal M}_{\lambda}^{(b)}$ and determining
${\cal M}_{\lambda}^{(c)}$ to the required order in $\omega$
from the gauge invariance condition (\ref{3.16}) we get
for the case of real photon emission, $k^{2} = 0$,
the following [see (3.27) of \cite{Lebiedowicz:2021byo}]
where we neglect gauge terms $\propto k_{\lambda}$:
\begin{eqnarray}
&&{\cal M}_{\lambda}(p_{1}', p_{2}', k, p_{a}, p_{b})
= e \bigg{\lbrace}
\bigg{[} \frac{p_{a \lambda}}{p_{a} \cdot k} 
- \frac{p'_{1 \lambda}}{p_{1}' \cdot k} 
\bigg{]} 
{\cal M}^{(\rm on)}(\nu, t) \nonumber\\
&& \quad -2 
\bigg{[} 
p_{a\lambda} \frac{p_{b} \cdot k}{p_{a} \cdot k} - p_{b\lambda}
\bigg{]}
\frac{\partial}{\partial \nu} {\cal M}^{(\rm on)}(\nu, t)
\nonumber\\
&& \quad -2 
\bigg{[}\frac{p_{a\lambda}}{p_{a} \cdot k} 
- \frac{p_{1\lambda}}{p_{1} \cdot k} \bigg{]} 
\bigg{[} (p_{a} - p_{1},k) - p_{a} \cdot l_{1} \bigg{]}
\nonumber \\
&& \qquad \times 
\frac{\partial}{\partial t} {\cal M}^{(\rm on)}(\nu, t) \bigg{\rbrace}
+ {\cal O}(\omega)\,, \nonumber \\
&&\nu = s - 2 m_{\pi}^{2}, \quad 
t = (p_{a} - p_{1})^{2} = (p_{b} - p_{2})^{2}\,.
\label{4.11}
\end{eqnarray}
Here ${\cal M}^{(\rm on)}(\nu, t)$ is the on-shell
$\pi^{-} \pi^{0} \to \pi^{-} \pi^{0}$ amplitude (\ref{3.6a}).
We can, for consistency, still expand 
(cf. Appendix~\ref{sec:appendixA})
\begin{eqnarray}
\frac{p'_{1 \lambda}}{p_{1}' \cdot k}
&=&
\frac{(p_{1} - l_{1})_{\lambda}}{(p_{1} - l_{1}, k)} \nonumber \\
&=& \frac{p_{1\lambda}}{p_{1} \cdot k} 
+ \frac{1}{(p_{1} \cdot k)^{2}}
\bigg{[}p_{1 \lambda}(l_{1} \cdot k)-l_{1 \lambda}(p_{1} \cdot k) \bigg{]} \nonumber \\
&&+ \,{\cal O}(\omega) \,.
\label{4.12}
\end{eqnarray}
In this way we get [see (A1) of \cite{Lebiedowicz:2021byo}]
\begin{eqnarray}
&&{\cal M}_{\lambda}(p_{1}', p_{2}', k, p_{a}, p_{b})
= e \bigg{\lbrace}
\bigg{[} \frac{p_{a \lambda}}{p_{a} \cdot k} 
- \frac{p_{1 \lambda}}{p_{1} \cdot k} 
\bigg{]} 
{\cal M}^{(\rm on)}(\nu, t) \nonumber\\
&& \quad - \frac{1}{(p_{1} \cdot k)^{2}}
\bigg{[}p_{1 \lambda}(l_{1} \cdot k)-l_{1 \lambda}(p_{1} \cdot k) \bigg{]}
{\cal M}^{(\rm on)}(\nu, t) \nonumber\\
&& \quad -2 
\bigg{[} 
p_{a\lambda} \frac{p_{b} \cdot k}{p_{a} \cdot k} - p_{b\lambda}
\bigg{]}
\frac{\partial}{\partial \nu} {\cal M}^{(\rm on)}(\nu, t)
\nonumber\\
&& \quad -2 
\bigg{[}\frac{p_{a\lambda}}{p_{a} \cdot k} 
- \frac{p_{1\lambda}}{p_{1} \cdot k} \bigg{]} 
\bigg{[} (p_{a} - p_{1},k) - p_{a} \cdot l_{1} \bigg{]}
\nonumber \\
&& \qquad \times 
\frac{\partial}{\partial t} {\cal M}^{(\rm on)}(\nu, t) \bigg{\rbrace}
+ {\cal O}(\omega)\,.
\label{4.13}
\end{eqnarray}
We leave it to the reader to insert in (\ref{4.13})
$l_{1}$, $l_{2}$ and $k$ from (\ref{15})--(\ref{4.6}) and
to convince himself that in this way we have given
the terms of order
$\omega^{-1}$ and $\omega^{0}$ 
in the Laurent expansion of the amplitude
${\cal M}_{\lambda}$ 
for the reaction (\ref{3.7})
for $\omega \to 0$.

The first term on the r.h.s. of Eq.~(\ref{4.13}) is
the pole term $\propto \omega^{-1}$ and this
is exactly Weinberg's soft-photon term.
He writes in \cite{Weinberg:1965nx}:
``Hence the effect of attaching one soft-photon line
to an arbitrary diagram is simply to supply an extra factor,
\begin{eqnarray}
\sum_{n} e_{n} \eta_{n} 
\frac{p_{n}^{\mu}}{p_{n} \cdot q - i \eta_{n} \varepsilon} \,,
\label{4.14}
\end{eqnarray}
the sum running over all external lines in the original diagram.''
Here $q$ is the photon four-momentum and
$\eta_{n} = +1$ for an outgoing charged particle,
$\eta_{n} = -1$ for an incoming charged particle.
In our work, \cite{Lebiedowicz:2021byo} and (\ref{4.13}) here, we have given
the \textit{next-to-leading term to Weinberg's pole term}
for the reaction (\ref{3.7}).

%--------------------------------------------------
\section{Soft photon theorem II}
\label{sec:5}
%--------------------------------------------------

Now we want to discuss Low's version of soft-photon theorem \cite{Low:1958sn}.
Of course, as a starting point he considers again the diagrams
(a), (b), and (c) of Fig.~\ref{fig:4} for ${\cal M}_{\lambda}$.
He also uses the generalized Ward identity which gave us
(\ref{4.9}) for $\Delta_{F} \widehat{\Gamma}\ul^{(\gamma \pi\pi)}$.
Considering only real photon emission we have then for the term
${\cal M}_{\lambda}^{(a)}$ [see Eq.~(2.11) of \cite{Low:1958sn}
but note that a different metric convention is used there]
\begin{eqnarray}
&&{\cal M}_{\lambda}^{(a)}(p_{1}', p_{2}', k, p_{a}, p_{b})
= e 
{\cal M}[ (p_{a}-k, p_{b}) + p_{1}' \cdot p_{2}',  \nonumber \\
&& \qquad (p_{b}-p_{2}')^{2}, m_{a}^{2} = (p_{a}-k)^{2}, m_{\pi}^{2}, m_{\pi}^{2}, m_{\pi}^{2} ]
\frac{p_{a \lambda}}{p_{a} \cdot k}
\,. \nonumber \\
\label{5.1}
\end{eqnarray}
Now Low expands ${\cal M}$ with respect to $k$
keeping $p_{1}'$ and $p_{2}'$ fixed.
That is, he only expands with respect to $k$ which is \textit{explicit}
in the parametrization chosen.
In this way we get
\begin{eqnarray}
&&{\cal M}_{\lambda}^{(a)}(p_{1}', p_{2}', k, p_{a}, p_{b})
= e \frac{p_{a \lambda}}{p_{a} \cdot k} \nonumber\\
&& \quad \times
\bigg{\lbrace}
{\cal M}^{(\rm on)}[p_{a} \cdot p_{b} + p_{1}' \cdot p_{2}', (p_{b}-p_{2}')^{2}]
\nonumber\\
&& \qquad - (p_{b} \cdot k) \frac{\partial}{\partial \nu} 
{\cal M}^{(\rm on)}[p_{a} \cdot p_{b} + p_{1}' \cdot p_{2}', (p_{b}-p_{2}')^{2}]
\nonumber\\
&& \qquad - 2 (p_{a} \cdot k) \frac{\partial}{\partial m_{a}^{2}}
{\cal M}[ p_{a} \cdot p_{b} + p_{1}' \cdot p_{2}',
(p_{b}-p_{2}')^{2}, 
\nonumber\\
&& \qquad \qquad
\left. m_{a}^{2}, m_{\pi}^{2}, m_{\pi}^{2}, m_{\pi}^{2} ]\right|_{m_{a}^{2} = m_{\pi}^{2}}
\bigg{\rbrace}
+ {\cal O}(k)\,.
\label{5.2}
\end{eqnarray}
Note an important point: whereas the expansion of the scalar function
${\cal M}$ on the r.h.s. of (\ref{5.1}) with respect to $k$,
keeping $p_{1}'$ and $p_{2}'$ fixed, is completely standard,
this is \textit{not the case} for ${\cal M}_{\lambda}^{(a)}$.
Keeping in ${\cal M}_{\lambda}^{(a)}(p_{1}', p_{2}', k, p_{a}, p_{b})$
$p_{a}$, $p_{b}$, $p_{1}'$, and $p_{2}'$ fixed and expanding in $k$,
in the usual sense with varying $k$,
we go \textit{outside} the physical region where we must have
$p_{a} + p_{b} = p'_{1} + p'_{2} + k$ (\ref{3.8}).
Thus, the r.h.s. of (\ref{5.2}) should,
in our opinion, be considered as giving an approximate expression for
${\cal M}_{\lambda}^{(a)}$ making sense \textit{only}
for the physical value of $k$ satisfying (\ref{3.8}).

Now we can treat ${\cal M}_{\lambda}^{(b)}$ in a similar way
and then determine ${\cal M}_{\lambda}^{(c)}$
approximately from the gauge-invariance condition (\ref{3.16}).
The result is Low's formula (see (1.7) of \cite{Low:1958sn})
\begin{eqnarray}
&&{\cal M}_{\lambda}(p_{1}', p_{2}', k, p_{a}, p_{b})
= e \bigg{\lbrace}
\bigg{[} \frac{p_{a \lambda}}{p_{a} \cdot k} 
- \frac{p'_{1 \lambda}}{p_{1}' \cdot k} 
\bigg{]} 
{\cal M}^{(\rm on)}(\nu_{L}, t_{2}) \nonumber\\
&& \quad - 
\bigg{[} 
p_{a\lambda} \frac{p_{b} \cdot k}{p_{a} \cdot k} 
+
p'_{1\lambda} \frac{p_{2}' \cdot k}{p_{1}' \cdot k} 
- p_{b\lambda} - p'_{2\lambda}
\bigg{]} \nn \\
&& \qquad \times
\frac{\partial}{\partial \nu_{L}} {\cal M}^{(\rm on)}(\nu_{L}, t_{2})
\bigg{\rbrace} + {\cal O}(k)\,,
\label{5.4}
\end{eqnarray}
where
\begin{eqnarray}
&&\nu_{L} = p_{a} \cdot p_{b} + p_{1}' \cdot p_{2}' 
=\nu - (p_{a}+p_{b},k) 
\,, \nonumber \\
&&t_{2} = (p_{b} - p_{2}')^{2} = (p_{a} - p_{1}' - k)^{2}\,.
\label{5.5}
\end{eqnarray}
Note that the amplitude ${\cal M}^{(\rm on)}$
is evaluated at values of the momenta 
%satisfying (\ref{3.8})
appropriate to the radiative process.
%In some works, e.g. Eq.~(17) of \cite{Liou:1987ug}, 
%Low's original result (\ref{5.4}) is rewritten with 
%$\nu_{L} \to \overline{s} = (s_{i} + s_{f})/2$,
%where $s_{i} = s = (p_{a} + p_{b})^{2}$, $s_{f} = (p_{1}' + p_{2}')^{2}$.
%But the correct relation is
%$\nu_{L} =  \overline{s} - 2 m_{\pi}^{2}$.

Again we emphasize that (\ref{5.4}) is \textit{not} the expansion of 
${\cal M}_{\lambda}$ \textit{around some phase-space point}.
The r.h.s. of (\ref{5.4}) gives an approximate expression for
${\cal M}_{\lambda}$ at a \textit{given phase-space point}
$p'_{1}, p'_{2}, k$.
Also, the leading approximation in (\ref{5.4}) does \textit{not} give
what is frequently called Low's theorem,
but really is Weinberg's version of the soft-photon theorem;
see (\ref{4.14}).
We see this best by considering the reactions
\begin{eqnarray}
\pi^{-} \,(p\ua)+ \pi^{+}\,(p\ub)&\to \pi^{-}\,(p\uu)+\pi^{+}\,(p\ud)\,, 
\label{5.6a}
\end{eqnarray}
and
\begin{eqnarray}
\pi^{-} \,(p\ua)+ \pi^{+}\,(p\ub)&\to \pi^{-}\,(p'\uu)+\pi^{+}\,(p'\ud)
+\gamma\,(k,\varepsilon)\,. \qquad
\label{5.6}
\end{eqnarray}
The leading term according to Low for (\ref{5.6}) is
\begin{eqnarray}
&&{\cal M}_{\lambda}(p_{1}', p_{2}', k, p_{a}, p_{b}) \nonumber \\
&& \quad = e \bigg{\lbrace}
\bigg{[} \frac{p_{a \lambda}}{p_{a} \cdot k} 
      - \frac{p'_{1 \lambda}}{p_{1}' \cdot k} 
\bigg{]} 
{\cal M}^{(\rm on)}(\nu_{L}, t_{2}) \nonumber\\
&& \qquad  + 
\bigg{[}- \frac{p_{b \lambda}}{p_{b} \cdot k} 
       + \frac{p'_{2 \lambda}}{p_{2}' \cdot k} 
\bigg{]}
{\cal M}^{(\rm on)}(\nu_{L}, t_{1}) \bigg{\rbrace}
+ {\cal O}(\omega^{0})\,, \qquad \;
\label{5.7}
\end{eqnarray}
where $t_{1}$ and $t_{2}$ are defined in (\ref{2.5}).
According to Weinberg, see (\ref{4.14}), we have,
on the other hand,
\begin{eqnarray}
&&{\cal M}_{\lambda}(p_{1}', p_{2}', k, p_{a}, p_{b})\nonumber \\
&& \quad = e \bigg{\lbrace}
\bigg{[} 
  \frac{p_{a \lambda}}{p_{a} \cdot k} 
- \frac{p_{1 \lambda}}{p_{1} \cdot k} 
- \frac{p_{b \lambda}}{p_{b} \cdot k} 
+ \frac{p_{2 \lambda}}{p_{2} \cdot k} 
\bigg{]}
{\cal M}^{(\rm on)}(\nu, t)
\bigg{\rbrace} \nonumber \\
&& \qquad + {\cal O}(\omega^{0})\,.
\label{5.8}
\end{eqnarray}
In (\ref{5.7}) we have an approximate expression for
${\cal M}_{\lambda}$ valid at the given phase-space point
$p_{1}'$, $p_{2}'$, $k$.
In (\ref{5.8}) we have the pole term of the Laurent expansion
of ${\cal M}_{\lambda}$
around the phase-space point 
$p_{1}' = p_{1}$, $p_{2}' = p_{2}$, $k = 0$.

Let us go back to the $\pi^{-} \pi^{0} \to \pi^{-} \pi^{0} \gamma$ case.
In (\ref{5.4}) we have Low's formula which gives us
an approximate expression for ${\cal M}_{\lambda}$
at a given phase-space point.
We can construct, as we did in Sec.~\ref{sec:4},
the corresponding expansion of this approximate expression
around the phase-space point $(k = 0, \bm{\hat{p}\uu})$.
Inserting in (\ref{5.4})
$p_{1}' = p_{1} - l_{1}$,
$p_{2}' = p_{2} - l_{2}$
from (\ref{4.2}) we get
\begin{eqnarray}
&&{\cal M}^{(\rm on)}(\nu_{L}, t_{2})\nn \\
&& \quad =
{\cal M}^{(\rm on)}[\nu - (p_{a}+p_{b},k), 
t - 2 \left( (p_{a} - p_{1},k) - p_{a} \cdot l_{1} \right) ]\nn \\
&&\qquad
+ \;{\cal O}(\omega^{2}) \nn \\
&&\quad = {\cal M}^{(\rm on)}(\nu, t)
- (p_{a} + p_{b},k) \frac{\partial}{\partial \nu} {\cal M}^{(\rm on)}(\nu, t)\nn \\
&& \qquad
- 2 \left( (p_{a} - p_{1},k) - p_{a} \cdot l_{1} \right) 
\frac{\partial}{\partial t} {\cal M}^{(\rm on)}(\nu, t)
+ {\cal O}(\omega^{2})\,.
\nn \\
\label{5.9}
\end{eqnarray}
Furthermore we find
\begin{eqnarray}
&&-\bigg{[} \frac{p_{a \lambda}}{p_{a} \cdot k} 
- \frac{p'_{1 \lambda}}{p_{1}' \cdot k} 
\bigg{]} (p_{a} + p_{b}, k)
\frac{\partial}{\partial \nu} {\cal M}^{(\rm on)}(\nu, t) \nonumber\\
&& - 
\bigg{[} 
p_{a\lambda} \frac{p_{b} \cdot k}{p_{a} \cdot k} 
+
p'_{1\lambda} \frac{p_{2}' \cdot k}{p_{1}' \cdot k} 
- p_{b\lambda} - p'_{2\lambda}
\bigg{]}
\frac{\partial}{\partial \nu_{L}} {\cal M}^{(\rm on)}(\nu_{L}, t_{2})\nonumber\\
&&=
-2 
\bigg{[} 
p_{a \lambda} \frac{p_{b} \cdot k}{p_{a} \cdot k} - p_{b \lambda}
\bigg{]} 
\frac{\partial}{\partial \nu} {\cal M}^{(\rm on)}(\nu, t)
+ \;{\cal O}(\omega) \,.
\label{5.9a}
\end{eqnarray}

From (\ref{5.4}), (\ref{5.9}) and (\ref{5.9a})
we get the result for ${\cal M}_{\lambda}$ identical to our result (\ref{4.11}).
We can go on to (\ref{4.13}) using (\ref{4.12}).
In this way we have given the relation between Low's theorem (\ref{5.4})
and the Laurent series (\ref{4.13}) where
the pole term $\propto \omega^{-1}$ is given by
Weinberg's soft-photon theorem and
the next-to-leading term $\propto \omega^{0}$ by
our calculation, (3.27) and (A1) of \cite{Lebiedowicz:2021byo}.

%--------------------------------------------------
\section{Cross section for $\pi^{-} \pi^{0} \to \pi^{-} \pi^{0} \gamma$}
\label{sec:6}
\begin{figure}[!ht]
\includegraphics[width=.38\textwidth]{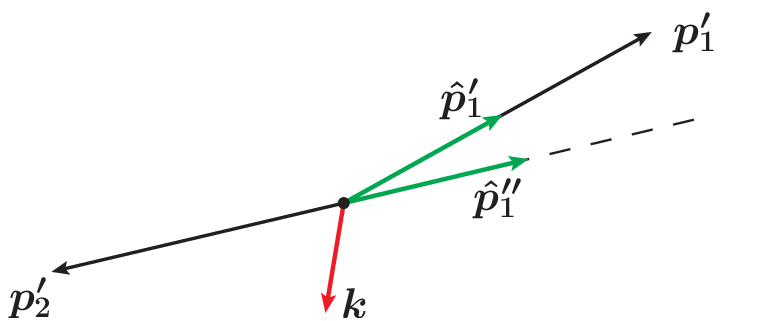}
\caption{Sketch of the momentum configuration for
the final state of
$\pi^{-}(p_{a}) \,\pi^{0}(p_{b}) 
\to \pi^{-}(p_{1}') \,\pi^{0}(p_{2}')\, \gamma(k)$ 
with the definition of the unit vectors 
$\bm{\hat{p}_{1}'}$ and 
$\bm{\hat{p}_{1}''}=-\bm{\hat{p}_{2}'}$.}
\label{fig:7}
\end{figure}

%--------------------------------------------------
Here we consider the cross section for $\pi^{-} \pi^{0}$ scattering
with real photon emission and summed over the photon polarizations.
We get
\bal{6.1}
&\dv \sigma (\pi^{-} (p\ua)+ \pi^{0}(p\ub) \to \pi^{-}(p'\uu)+\pi^{0}(p'\ud)+\gamma(k) ) \nn\\
&\quad =\frac{1}{2 \sqrt{s(s-4m_{\pi}^{2})}}
\frac{\dv^{3}k}{(2\pi)^{3}2k^{0}}\,
\frac{\dv^{3}p'\uu}{(2\pi)^{3}2p\uu^{\prime 0}}\,
\frac{\dv^{3}p'\ud}{(2\pi)^{3}2p\ud^{\prime 0}}\nn\\
&\qquad \times (2\pi)^{4}\delta^{(4)}(p'\uu+p'\ud+k-p\ua-p\ub)\nn\\
&\qquad \times (-1) {\cal M}_{\lambda}(p'\uu,p'\ud,k,p\ua,p\ub)
{\cal M}^{\lambda *}(p'\uu,p'\ud,k,p\ua,p\ub) 
\end{align}
with ${\cal M}_{\lambda}$ from (\ref{3.9}).
We are interested here in small $\omega$
where we found that the phase space can be parametrized by $(k, \bm{\hat{p}\uu'})$;
see (\ref{2.7}).
Here we have for real photons, of course,
\begin{align}
\label{4.5_new}
(k^{\mu})&=
\omega \left( \begin{array}{l}
1
\vspace{0.1cm}\\
\bm{\hat{k}}
\end{array}\right), \quad 
|\bm{\hat{k}}| = 1\,.
\end{align}
We can, as well, choose $(k, \bm{\hat{p}_{2}'})$
as phase-space variables. 
Below we shall discuss the following cross sections
\bal{6.2a}
\sigma_{1} = \frac{\omega \,\dv\sigma}{\dv \omega \, 
\dv \Omega_{\hat{k}} \, \dv \Omega_{\hat{p}_{1}'}}\,, \quad 
\sigma_{2} = \frac{\omega \,\dv\sigma}{\dv \omega \, 
\dv \Omega_{\hat{k}} \, \dv \Omega_{\hat{p}_{2}'}}\,, \quad 
\sigma_{3} = \frac{\omega \,\dv\sigma}{\dv \omega}\,,
\end{align}
where $\dv \Omega_{\hat{k}}$, $\dv \Omega_{\hat{p}_{1}'}$,
and $\dv \Omega_{\hat{p}_{2}'}$ are the solid-angle elements
to $\bm{\hat{k}}$, $\bm{\hat{p}_{1}'}$, and $\bm{\hat{p}_{2}'}$ 
in the overall c.m. system, respectively.
For calculating the expansions in $\omega$
of the cross sections (\ref{6.2a}) it is important to choose
the appropriate expansion of ${\cal M}_{\lambda}(p'\uu,p'\ud,k,p\ua,p\ub)$.
For calculating the cross section with respect to 
$(\bm{\hat{k}}, \bm{\hat{p}_{1}'})$ 
we shall use the expansion around
$(\bm{\hat{k}} = 0, \bm{\hat{p}_{1}'})$
keeping $\bm{\hat{p}_{1}'}$ constant.
Similarly, for the cross section with respect to $\bm{\hat{k}}$
and $\bm{\hat{p}_{2}'}$ it will be convenient to use the expansion where $\bm{\hat{p}_{2}'}$ is kept constant.
We illustrate this in Fig.~\ref{fig:7}.
We set
\begin{align}
\label{6.2a1}
\bm{\hat{p}_{1}''} = - \bm{\hat{p}_{2}'}
\end{align}
and get, after a simple calculation,
\begin{align}
\label{6.2b}
\bm{\hat{p}_{1}'} = \bm{\hat{p}_{1}''}
- \frac{\omega}{|\bm{p_{1}}|}
\left[ 
\bm{\hat{k}} (\bm{\hat{p}_{1}''})^{2} - 
(\bm{\hat{p}_{1}''} \cdot \bm{\hat{k}})
\bm{\hat{p}_{1}''}
\right] 
+ {\cal O}(\omega^{2})\,.
\end{align}
%
%---------------------------------------------------------
%---------------------------------------------------------

\begin{figure}[!ht]
\includegraphics[width=.25\textwidth]{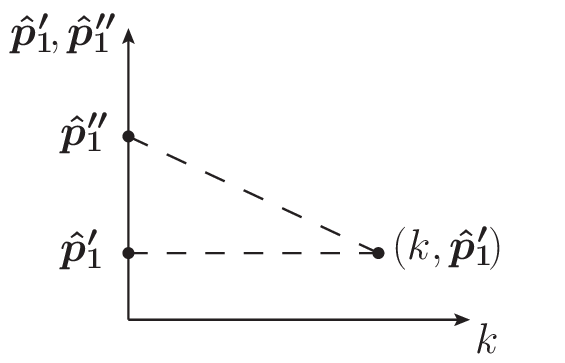}
\caption{Sketch for the two expansions (\ref{6.2c}) of
${\cal M}_{\lambda}(p'\uu,p'\ud,k,p\ua,p\ub)$
in the phase space $(k, \bm{\hat{p}_{1}'})$.}
\label{fig:8}
\end{figure}

The next point to realize is, that in expanding
${\cal M}_{\lambda}(p'\uu,p'\ud,k,p\ua,p\ub)$
we have the freedom to choose the starting point
of the expansion appropriately, of course,
always staying with $(p_{1}, p_{2})$
close to $(p_{1}', p_{2}')$
as sketched in Fig.~\ref{fig:6}.
Referring always to the phase-space variables
$(k, \bm{\hat{p}_{1}'})$ we choose as a starting point for
expanding $\sigma_{1}$ from (\ref{6.2a})
the point $(k = 0, \bm{\hat{p}_{1}'})$,
for $\sigma_{2}$ the point $(k = 0, \bm{\hat{p}_{1}''})$.
Then, $\sigma_{3}$ from (\ref{6.2a}) 
should be independent of these two choices
and this is indeed what we shall see below.
We can, therefore, write the following expansions:
\begin{eqnarray}
&&\omega {\cal M}_{\lambda}(p'\uu,p'\ud,k,p\ua,p\ub) \nn\\
&&= 
\widehat{\cal M}_{\lambda}^{(0)}
(s,\bm{\hat{p}\ua},\bm{\hat{p}\uu'},\bm{\hat{k}})
+
\omega
\widehat{\cal M}_{\lambda}^{(1)}
(s,\bm{\hat{p}\ua},\bm{\hat{p}\uu'},\bm{\hat{k}};\bm{\hat{p}\uu'})
+
{\cal O}(\omega^{2}) \nn\\
&&= 
\widehat{\cal M}_{\lambda}^{(0)}
(s,\bm{\hat{p}\ua},\bm{\hat{p}\uu''},\bm{\hat{k}})
+
\omega
\widehat{\cal M}_{\lambda}^{(1)}
(s,\bm{\hat{p}\ua},\bm{\hat{p}\uu'},\bm{\hat{k}};\bm{\hat{p}\uu''})
+
{\cal O}(\omega^{2})\,.\nn\\
\label{6.2c}
\end{eqnarray}
Here we indicate by the last variable in $\widehat{\cal M}_{\lambda}^{(1)}$
the starting point of the expansion,
$(k = 0, \bm{\hat{p}_{1}'})$ or $(k = 0, \bm{\hat{p}_{1}''})$,
respectively; see Fig.~\ref{fig:8}.
The precise definitions of $\widehat{\cal M}_{\lambda}^{(0)}$
and $\widehat{\cal M}_{\lambda}^{(1)}$ following from
(\ref{4.13}) are given below.
%--------------------------------------------------------------
%--------------------------------------------------------------

Now we come to the calculation of the expansion of $\sigma_{1}$ (\ref{6.2a}).

The cross section with respect to the phase-space variables 
$\omega$, $\bm{\hat{k}}$, $\bm{\hat{p}\uu'}$
reads (see Appendix~B of \cite{Lebiedowicz:2021byo})
\bal{6.3}
&\dv\sigma(\pi^{-} \pi^{0}\to \pi^{-} \pi^{0} \gamma)
=
\frac{1}{\sqrt{s(s-4m_{\pi}^{2})}}
\frac{1}{2^{4}(2\pi)^{5}} \nn\\
& \quad \times \omega \, \dv\omega \, \dv \Omega_{\hat{k}} \, \dv \Omega_{\hat{p}_{1}'} \, 
J(s, \omega, \bm{\hat{p}\uu'}, \bm{\hat{k}})\,
(- {\cal M}_{\lambda} {\cal M}^{\lambda *})\,.
\end{align}
Here $J$ is a kinematic function given in
(B7) of \cite{Lebiedowicz:2021byo}
and we consider ${\cal M}_{\lambda}$ as a function
of the independent initial variables $s$,
$\bm{\hat{p}\ua}$ and the phase-space variables (\ref{2.7})
\bal{6.4}
{\cal M}_{\lambda}(p'\uu,p'\ud,k,p\ua,p\ub) \equiv {\cal M}_{\lambda}
(s,\bm{\hat{p}\ua},\omega,\bm{\hat{k}},\bm{\hat{p}\uu'})\,.
\end{align}

We are interested in the cross section (\ref{6.3})
for \mbox{$\omega \to 0$}.
The expansion of the phase-space factor $J$
is easily obtained from (B3)--(B8) of \cite{Lebiedowicz:2021byo}
\bal{6.5}
&J(s, \omega, \bm{\hat{p}\uu'}, \bm{\hat{k}}) 
= J^{(1)}(s, \omega, \bm{\hat{p}\uu'}, \bm{\hat{k}})
+ {\cal O}(\omega^{2})\,,
\nn\\
& J^{(1)}(s, \omega, \bm{\hat{p}\uu'}, \bm{\hat{k}}) \nn \\
& \quad = 
\frac{1}{2} \sqrt{1 - \frac{4 m_{\pi}^{2}}{s}}
- \frac{\omega}{\sqrt{s}}
\bigg{(}
\frac{2 m_{\pi}^{2}}{\sqrt{s(s-4 m_{\pi}^{2})}} + \bm{\hat{p}\uu'} \cdot \bm{\hat{k}}
\bigg{)} \,.
\end{align}

The expansion in $\omega$ of ${\cal M}_{\lambda}$
(\ref{6.4}) was the topic of Sec.~\ref{sec:4}.
Now we consider given values for $\bm{\hat{k}}$
and $\bm{\hat{p}\uu'}$ and vary $\omega$.
Therefore, in the schematic diagrams of Fig.~\ref{fig:5} 
and Fig.~\ref{fig:8}
we move
for fixed $\bm{\hat{p}\uu'} = \bm{\hat{p}\uu}$,
that is, for fixed $\bm{l_{1 \perp}} = 0$
[see (\ref{4.1})] along the line 
$k = \omega (1, \bm{\hat{k}})^{\top}$.
\mbox{The expansion} of ${\cal M}_{\lambda}$ (\ref{6.4})
on this line is given in (\ref{4.13}), of course,
inserting $l_{1}$ and $l_{2}$ from (\ref{15}) 
with $\bm{l_{1 \perp}} = 0$.
We denote the corresponding values of $l_{i}$ by $l_{i}'$
($i = 1, 2$) in the following.
In this way we obtain with 
$\bm{\hat{p}\uu'} = \bm{\hat{p}\uu}$
[see (B12)--(B14) of \cite{Lebiedowicz:2021byo}]
for the first line on the r.h.s. of (\ref{6.2c})
\begin{eqnarray}
{\cal M}_{\lambda}(s,\bm{\hat{p}\ua},\omega,\bm{\hat{k}},\bm{\hat{p}\uu'}) =
\frac{1}{\omega}\widehat{{\cal M}}_{\lambda}^{(0,1)}+
                \widehat{{\cal M}}_{\lambda}^{(1,1)}+
{\cal O}(\omega)\,, \qquad \quad
\label{6.6}
\end{eqnarray}
where
\begin{eqnarray}
\widehat{{\cal M}}_{\lambda}^{(0,1)} &=&
\widehat{{\cal M}}_{\lambda}^{(0)}(s, \bm{\hat{p}\ua}, \bm{\hat{p}\uu'}, \bm{\hat{k}}) \nonumber\\
&=& e \omega \bigg{[} 
       \frac{p_{a \lambda}}{p_{a} \cdot k} 
-      \frac{p_{1 \lambda}}{p_{1} \cdot k} \bigg{]}
{\cal M}^{(\rm on)}(\nu, t)\,,
\label{6.7}\\
\widehat{{\cal M}}_{\lambda}^{(1,1)} &=&
\widehat{{\cal M}}_{\lambda}^{(1)}(s, \bm{\hat{p}\ua}, \bm{\hat{p}\uu'}, \bm{\hat{k}}; \bm{\hat{p}\uu'}) \nonumber\\
&=& e 
\bigg{\lbrace}
-\frac{1}{(p_{1} \cdot k)^{2}}
\bigg{[} p_{1 \lambda} (l_{1}' \cdot k) - l_{1 \lambda}' (p_{1} \cdot k) \bigg{]} 
\nonumber\\ 
&&
\times {\cal M}^{(\rm on)}(\nu, t) \nonumber\\ 
&&
- 2 \bigg{[} p_{a\lambda} \frac{p_{b} \cdot k}{p_{a} \cdot k} - p_{b\lambda} \bigg{]}
\frac{\partial}{\partial \nu}  {\cal M}^{(\rm on)}(\nu, t)\nonumber \\
&&
- 2 \bigg{[} \frac{p_{a\lambda}}{p_{a} \cdot k} - \frac{p_{1\lambda}}{p_{1} \cdot k} \bigg{]}
\bigg{[} (p_{a} - p_{1},k) - (p_{a} \cdot l_{1}') \bigg{]} 
\nonumber\\
&& \times
\frac{\partial}{\partial t} {\cal M}^{(\rm on)}(\nu, t)
\bigg{\rbrace}\,.
\label{6.8}
\end{eqnarray}
Note that both $\widehat{{\cal M}}_{\lambda}^{(0,1)}$
and $\widehat{{\cal M}}_{\lambda}^{(1,1)}$ are independent of $\omega$
and they are unambiguously defined inserting 
$l_{1}'$ and $l_{2}'$ which are the values of
$l_{1}$ and $l_{2}$
from (\ref{15}) with $\bm{l_{1 \perp}} = 0$
and $\nu = s - 2 m_{\pi}^{2}$, $t = (p_{a} - p_{1})^{2}$
from (\ref{3.4})--(\ref{3.6}).
Note also that the expansion (\ref{4.12}) 
which is used here is,
for our case $\bm{l_{1 \perp}} = 0$, alright for
$\omega \ll |\bm{p\uu}|$ and
all values of $\bm{\hat{p}\uu}$ and $\bm{\hat{k}}$;
see (\ref{A3}).

Now we define
\begin{eqnarray}
&&{\cal A}^{(0,1)}(s, \bm{\hat{p}\ua}, \bm{\hat{p}\uu'}, \bm{\hat{k}}) =
\frac{1}{\sqrt{s(s-4m_{\pi}^{2})}}
\frac{1}{2^{4}(2\pi)^{5}} \nonumber\\
&& \quad \times
\frac{1}{2} \sqrt{1 - \frac{4 m_{\pi}^{2}}{s}}
\left( - \widehat{{\cal M}}_{\lambda}^{(0,1)} \widehat{{\cal M}}^{(0,1)\lambda *} \right)\,,
\label{6.9}\\
&&{\cal A}^{(1,1)}(s, \bm{\hat{p}\ua}, \bm{\hat{p}\uu'}, \bm{\hat{k}}; \bm{\hat{p}\uu'}) =
\frac{1}{\sqrt{s(s-4m_{\pi}^{2})}}
\frac{1}{2^{4}(2\pi)^{5}} \nonumber\\
&& \quad \times
\bigg{[} \frac{1}{2} \sqrt{1 - \frac{4 m_{\pi}^{2}}{s}}
\left( - \widehat{{\cal M}}_{\lambda}^{(1,1)} \widehat{{\cal M}}^{(0,1)\lambda *} 
       - \widehat{{\cal M}}_{\lambda}^{(0,1)} \widehat{{\cal M}}^{(1,1)\lambda *} \right)
       \nonumber\\
&& \qquad - 
\frac{1}{\sqrt{s}}
\bigg{(}
\frac{2 m_{\pi}^{2}}{\sqrt{s(s-4 m_{\pi}^{2})}} 
+ \bm{\hat{p}\uu'} \cdot \bm{\hat{k}}
\bigg{)}
\left( - \widehat{{\cal M}}_{\lambda}^{(0,1)} \widehat{{\cal M}}^{(0,1)\lambda *} \right)
\bigg{]}
\,. \nonumber\\
\label{6.10}
\end{eqnarray}
Inserting (\ref{6.5})--(\ref{6.8}) in (\ref{6.3})
and using (\ref{6.9}) and (\ref{6.10}) we get
\begin{eqnarray}
&&\frac{\omega \,\dv\sigma(\pi^{-} \pi^{0}\to \pi^{-} \pi^{0} \gamma)}{\dv\omega \, \dv \Omega_{\hat{k}} \, \dv \Omega_{\hat{p}_{1}'}} \nn\\
&&\quad =\frac{1}{\sqrt{s(s-4m_{\pi}^{2})}}
\frac{1}{2^{4}(2\pi)^{5}}
J^{(1)}(s, \omega, \bm{\hat{p}\uu'}, \bm{\hat{k}}) \nn \\
&& \qquad  \times (-1) 
\left( \widehat{{\cal M}}_{\lambda}^{(0,1)} + \omega \widehat{{\cal M}}_{\lambda}^{(1,1)} \right)
\left( \widehat{{\cal M}}^{(0,1)\lambda}    + \omega \widehat{{\cal M}}^{(1,1)\lambda} \right)^{*} \nn\\
&& \qquad  +
{\cal O}(\omega^{2})
\nn \\
&&\quad ={\cal A}^{(0,1)}(s, \bm{\hat{p}\ua}, \bm{\hat{p}\uu'}, \bm{\hat{k}}) 
+ \omega {\cal A}^{(1,1)}(s, \bm{\hat{p}\ua}, \bm{\hat{p}\uu'}, \bm{\hat{k}}; \bm{\hat{p}\uu'})\nn \\
&&
\qquad 
+
{\cal O}(\omega^{2})\,.
\label{6.11}
\end{eqnarray}
Integrating (\ref{6.11}) over the solid angles of
$\bm{\hat{k}}$ and $\bm{\hat{p}\uu'}$
we find the expansion of the cross section
$\omega \, \dv\sigma / \dv\omega$ for $\omega \to 0$
\begin{eqnarray}
&&\omega \frac{\dv\sigma}{\dv\omega} (\pi^{-} \pi^{0}\to \pi^{-} \pi^{0} \gamma) \nn\\
&&\quad =\frac{1}{\sqrt{s(s-4m_{\pi}^{2})}}
\frac{1}{2^{4}(2\pi)^{5}}
\int
\dv \Omega_{\hat{k}} \, \dv \Omega_{\hat{p}_{1}'}
J^{(1)}(s, \omega, \bm{\hat{p}\uu'}, \bm{\hat{k}}) \nn \\
&& \qquad  \times (-1) 
\left( \widehat{{\cal M}}_{\lambda}^{(0,1)} + \omega \widehat{{\cal M}}_{\lambda}^{(1,1)} \right)
\left( \widehat{{\cal M}}^{(0,1)\lambda}    + \omega \widehat{{\cal M}}^{(1,1)\lambda} \right)^{*} \nn\\
&& \qquad  +
{\cal O}(\omega^{2})
\nn \\
&& \quad = \int
\dv \Omega_{\hat{k}} \, \dv \Omega_{\hat{p}_{1}'}
 {\cal A}^{(0,1)}(s, \bm{\hat{p}\ua}, \bm{\hat{p}\uu'}, \bm{\hat{k}}) \nn \\
&& \qquad
+ \, \omega \int
\dv \Omega_{\hat{k}} \, \dv \Omega_{\hat{p}_{1}'}
{\cal A}^{(1,1)}(s, \bm{\hat{p}\ua}, \bm{\hat{p}\uu'}, \bm{\hat{k}}; \bm{\hat{p}\uu'})
+
{\cal O}(\omega^{2})\,. \nn
\\
\label{6.12}
\end{eqnarray}
Note that in the expansions (\ref{6.11}) and (\ref{6.12})
all terms are unambiguously defined.
The expansion coefficients are independent of $\omega$,
as it should be.

Next we consider the cross section $\sigma_{2}$ from
(\ref{6.2a}), that is, the cross section with respect to 
$\omega$, $\bm{\hat{k}}$, and $\bm{\hat{p}_{2}'}$.
%Similarly to (\ref{6.7}) 
For this
we define here 
[see (\ref{6.2a1})--(\ref{6.2c})] 
the pion momenta $p_{1}''$ and $p_{2}''$
of the now appropriate nonradiative starting point
of the expansion (see Figs.~\ref{fig:7} and \ref{fig:8}).
We have
\begin{eqnarray}
\bm{p_{1}''} &=& 
- \bm{p_{2}''} = \bm{\hat{p}_{1}''} \,|\bm{p_{1}}|
= \bm{\hat{p}_{1}''} \sqrt{ \frac{s}{4} - m_{\pi}^{2}}\,,\nn\\
p_{1}''^{0} &=& p_{2}''^{0} = \frac{\sqrt{s}}{2}\,, \nn\\
t'' &=& (p_{a} - p_{1}'')^{2}\,.
\label{6.12a}
\end{eqnarray}
We get then
the following matrix elements 
for the second line on the r.h.s. of (\ref{6.2c})
using (\ref{4.13}):
\begin{eqnarray}
\widehat{{\cal M}}_{\lambda}^{(0,2)} &=&
\widehat{{\cal M}}_{\lambda}^{(0)}(s, \bm{\hat{p}\ua}, \bm{\hat{p}\uu''}, \bm{\hat{k}}) \nonumber\\
&=& e \omega \bigg{[} 
       \frac{p_{a \lambda}}{p_{a} \cdot k} 
-      \frac{p_{1 \lambda}''}{p_{1}'' \cdot k} \bigg{]}
{\cal M}^{(\rm on)}(\nu, t'')\,,
\label{6.13}\\
\widehat{{\cal M}}_{\lambda}^{(1,2)} &=&
\widehat{{\cal M}}_{\lambda}^{(1)}(s, \bm{\hat{p}\ua}, \bm{\hat{p}\uu'}, \bm{\hat{k}}; \bm{\hat{p}\uu''}) \nonumber\\
&=& e 
\bigg{\lbrace}
-\frac{1}{(p_{1}'' \cdot k)^{2}}
\bigg{[} p_{1 \lambda}'' (l_{1}'' \cdot k) 
- l_{1 \lambda}'' (p_{1}'' \cdot k) \bigg{]} \nonumber \\
&& \times
{\cal M}^{(\rm on)}(\nu, t'') \nonumber\\ 
&&
- 2 \bigg{[} p_{a\lambda} \frac{p_{b} \cdot k}{p_{a} \cdot k} - p_{b\lambda} \bigg{]}
\frac{\partial}{\partial \nu}  {\cal M}^{(\rm on)}(\nu, t'')\nonumber \\
&&
- 2 \bigg{[} \frac{p_{a\lambda}}{p_{a} \cdot k} 
- \frac{p_{1\lambda}''}{p_{1}'' \cdot k} \bigg{]}
\bigg{[} (p_{a} - p_{1}'',k) - (p_{a} \cdot l_{1}'') \bigg{]} 
\nonumber\\
&& \times
\frac{\partial}{\partial t''} {\cal M}^{(\rm on)}(\nu, t'')
\bigg{\rbrace} \,.
\label{6.14}
\end{eqnarray}
Here we have to set in (\ref{4.13}) $l_{1} = l_{1}''$
and $l_{2} = l_{2}''$ according to (\ref{15})
but with the replacements
\begin{eqnarray}
&&p_{1} \to p_{1}'', \quad \bm{\hat{p}_{1}} \to \bm{\hat{p}_{1}''},
\quad p_{2} \to p_{2}'', \nn \\
&&\bm{l_{1 \perp}} \to \bm{l_{1 \perp}''} = \bm{k} (\bm{\hat{p}_{1}''})^{2}
- (\bm{k} \cdot \bm{\hat{p}_{1}''}) \bm{\hat{p}_{1}''} \,.
\label{6.15}
\end{eqnarray}
We get then $\bm{l_{2 \perp}''} = 0$, 
that is,
$\bm{\hat{p}_{2}'} = \bm{\hat{p}_{2}''}$,
as we require here.

For the cross section $\sigma_{2}$ of (\ref{6.2a}) we obtain
\begin{eqnarray}
&&\frac{\omega \,\dv\sigma(\pi^{-} \pi^{0}\to \pi^{-} \pi^{0} \gamma)}{\dv\omega \, \dv \Omega_{\hat{k}} \, \dv \Omega_{\hat{p}_{2}'}} \nn\\
&&\quad =\frac{1}{\sqrt{s(s-4m_{\pi}^{2})}}
\frac{1}{2^{4}(2\pi)^{5}}
J^{(1)}(s, \omega, \bm{\hat{p}_{2}'}, \bm{\hat{k}}) \nn \\
&& \qquad  \times (-1) 
\left( \widehat{{\cal M}}_{\lambda}^{(0,2)} + \omega \widehat{{\cal M}}_{\lambda}^{(1,2)} \right)
\left( \widehat{{\cal M}}^{(0,2)\lambda}    + \omega \widehat{{\cal M}}^{(1,2)\lambda} \right)^{*} \nn\\
&& \qquad  +
{\cal O}(\omega^{2})
\nn \\
&&\quad =\frac{1}{\sqrt{s(s-4m_{\pi}^{2})}}
\frac{1}{2^{4}(2\pi)^{5}}
\nn \\
&&\qquad \times
\bigg{[}
\frac{1}{2} \sqrt{1 - \frac{4 m_{\pi}^{2}}{s}}
- \frac{\omega}{\sqrt{s}}
\bigg{(}
\frac{2 m_{\pi}^{2}}{\sqrt{s(s-4 m_{\pi}^{2})}} + \bm{\hat{p}\ud'} \cdot \bm{\hat{k}}
\bigg{)}
\bigg{]}
\nn \\
&& \qquad  \times (-1) 
\left( \widehat{{\cal M}}_{\lambda}^{(0,2)} + \omega \widehat{{\cal M}}_{\lambda}^{(1,2)} \right)
\left( \widehat{{\cal M}}^{(0,2)\lambda}    + \omega \widehat{{\cal M}}^{(1,2)\lambda} \right)^{*}
\nn \\
&&
\qquad 
+
{\cal O}(\omega^{2})\,,
\label{6.16}
\end{eqnarray}
and finally
\begin{eqnarray}
&&\frac{\omega \,\dv\sigma(\pi^{-} \pi^{0}\to \pi^{-} \pi^{0} \gamma)}{\dv\omega \, \dv \Omega_{\hat{k}} \, \dv \Omega_{\hat{p}_{2}'}} =
\frac{1}{\sqrt{s(s-4m_{\pi}^{2})}}
\frac{1}{2^{4}(2\pi)^{5}} \nonumber\\
&& \; \times
\bigg{\lbrace} \frac{1}{2} \sqrt{1 - \frac{4 m_{\pi}^{2}}{s}}
\left( - \widehat{{\cal M}}_{\lambda}^{(0,2)}                                     
         \widehat{{\cal M}}^{(0,2)\lambda *} 
\right)
\nonumber\\
&& \; +
\omega \bigg{[} \frac{1}{2} \sqrt{1 - \frac{4 m_{\pi}^{2}}{s}}
\left( - \widehat{{\cal M}}_{\lambda}^{(1,2)} \widehat{{\cal M}}^{(0,2)\lambda *} 
       - \widehat{{\cal M}}_{\lambda}^{(0,2)} \widehat{{\cal M}}^{(1,2)\lambda *} \right)
       \nonumber\\
&& \; - 
\frac{1}{\sqrt{s}}
\bigg{(}
\frac{2 m_{\pi}^{2}}{\sqrt{s(s-4 m_{\pi}^{2})}} 
+ \bm{\hat{p}\ud'} \cdot \bm{\hat{k}}
\bigg{)}
\left( - \widehat{{\cal M}}_{\lambda}^{(0,2)} \widehat{{\cal M}}^{(0,2)\lambda *} \right)
\bigg{]} \bigg{\rbrace}
\nn \\
&&
\; 
+
{\cal O}(\omega^{2})\,.
\label{6.16a}
\end{eqnarray}

Now we turn to $\sigma_{3}$ in (\ref{6.2a}).
In (\ref{6.12}) we have obtained 
$\omega \,\dv\sigma / \dv\omega$ by integrating
over the solid angles of $\bm{\hat{k}}$
and~$\bm{\hat{p}_{1}'}$.
We can, however, also integrate (\ref{6.16})
over the solid angles of $\bm{\hat{k}}$
and $\bm{\hat{p}_{2}'}$.
Will we get the same result up to order $\omega$?
From (\ref{6.16}) we get
\begin{eqnarray}
&&\omega \frac{\dv\sigma}{\dv\omega} (\pi^{-} \pi^{0}\to \pi^{-} \pi^{0} \gamma) \nn\\
&&\quad =\frac{1}{\sqrt{s(s-4m_{\pi}^{2})}}
\frac{1}{2^{4}(2\pi)^{5}}
\int
\dv \Omega_{\hat{k}} \, \dv \Omega_{\hat{p}_{2}'}
J^{(1)}(s, \omega, \bm{\hat{p}_{2}'}, \bm{\hat{k}}) \nn \\
&& \qquad  \times (-1) 
\left( \widehat{{\cal M}}_{\lambda}^{(0,2)} + \omega \widehat{{\cal M}}_{\lambda}^{(1,2)} \right)
\left( \widehat{{\cal M}}^{(0,2)\lambda}    + \omega \widehat{{\cal M}}^{(1,2)\lambda} \right)^{*} \nn\\
&& \qquad  +
{\cal O}(\omega^{2})\,.
\label{6.17}
\end{eqnarray}
In Appendix~\ref{sec:appendixB} we give the result for
the change of measure under the variable transformation
$\bm{\hat{p}_{1}'} \to \bm{\hat{p}_{1}''} = - \bm{\hat{p}_{2}'}$
from (\ref{6.2b}) for fixed $\bm{\hat{k}}$.
We find from (\ref{B9})
\begin{eqnarray}
&&\dv \Omega_{\hat{p}_{1}'}
J^{(1)}(s, \omega, \bm{\hat{p}_{1}'}, \bm{\hat{k}})\nn\\
&&\quad = \dv \Omega_{\hat{p}_{1}''}
J^{(1)}(s, \omega, -\bm{\hat{p}_{1}''}, \bm{\hat{k}}) 
+ {\cal O}(\omega^{2})\nn\\
&&\quad = \dv \Omega_{\hat{p}_{2}'}
J^{(1)}(s, \omega, \bm{\hat{p}_{2}'}, \bm{\hat{k}}) 
+ {\cal O}(\omega^{2})\,.
\label{6.18}
\end{eqnarray}
Inserting (\ref{6.18}) in (\ref{6.17}) and using (\ref{6.2c}), (\ref{6.6})--(\ref{6.8}),
(\ref{6.13}), and (\ref{6.14}), we find, indeed, 
that the expansions of the cross sections
$\omega \,\dv\sigma / \dv\omega$ from
(\ref{6.12}) and (\ref{6.17}) 
are the same up to order $\omega$
which is the order up to which we calculate here.

To conclude this chapter, we emphasize that for
the discussions of cross sections in (\ref{6.2a})
it was essential to have at our disposal
the general expansion of the radiative amplitude
${\cal M}_{\lambda}$ around a phase-space point
$(k = 0, \bm{\hat{p}_{1}})$, respectively
$(k = 0, \bm{l_{1 \perp}} = 0)$;
see Fig.~\ref{fig:5}.
The expansion parameters were $(k, \bm{l_{1 \perp}})$.
We found that in calculating $\sigma_{1}$ and $\sigma_{2}$
of (\ref{6.2a}) we had to use the general expansion
(\ref{4.13}) of ${\cal M}_{\lambda}$ but with \textit{different}
starting points and \textit{different} 
values of $\bm{l_{1 \perp}}$,
respectively; see Fig.~\ref{fig:8}.

%--------------------------------------------------
\section{Outline of the calculation for $\pi p \to \pi p \gamma$}
\label{sec:7}
%--------------------------------------------------
We use the framework of QCD and treat electromagnetism to lowest relevant order.
In QCD we have the symmetries: parity ($P$), charge conjugation ($C$), 
and time reversal ($T$).
These give us restrictions for the \mbox{propagators}, vertices, and amplitudes. 
Furthermore, we use the generalized Ward identities for pions and the proton 
\cite{Ward:1950xp,Takahashi:1957xn} 
and the Landau conditions for determining 
the singularities in amplitudes \cite{Landau:1959fi,Bjorken:1965}.
All our results are derived using only these rigorous methods.

Consider now the reactions \er{1} where energy-momentum conservation reads 
\bel{3}
p\ua+p\ub=p\uu+p\ud\,.
\ee
Thus, there are only three independent momenta which we choose as
\bal{4}
p\us &=p\ua + p\ub = p\uu +p\ud\,,\nn\\
p\ut &=p\ua - p\uu = p\ud -p\ub\,,\nn\\
p_{u} &=p\ua - p\ud = p\uu - p\ub\,.
\end{align}
We have
\bel{5}
s=p\us\2\;,\quad t=p\ut\2 \;,\quad u=p_{u}\2\,.
\ee
The amplitude for \er{1} has the general structure
\bal{6}
&\braket{\ppm (p\uu), \,  p(p\ud ,\lambda\ud)|{\mathcal T}|\ppm(p\ua), \, p(p\ub ,\lambda\ub)} \nn\\
& \quad = \bar{u}(p\ud , \lambda\ud)\bigg{[}A^{(\text{on})\pm}(s,t)\nn\\
& \qquad 
+\frac{1}{2}(\slash{p}_{a}+\slash{p}_{1})
B^{(\text{on})\pm}(s,t)\bigg{]}u(p\ub,\lambda\ub )\,,
\end{align}
with invariant functions $A^{(\text{on})\pm}$ and $B^{(\text{on})\pm}$; see e.g. \cite{Bjorken:1965}. 
In the calculation of the amplitude for \er{2} we need, however, the off-shell amplitude for $\pi p \to \pi p$ which is much more complicated than \er{6}. 
Writing for the on- or off-shell momenta of the general reaction \er{1} 
$\tp\ua$, $\tp\ub$, $\tp\uu$, $\tp\ud$ and defining $\tp\us$, $\tp\ut$, $\tp_{u}, \tilde{s}$, $\tilde{t}$ in analogy to \er{4} and \er{5} 
we find for the off-shell amplitudes 
\begin{align}
\label{7}
&\cM^{(0)\pm}(\tp\uu ,\,\tp\ud ,\, \tp\ua ,\,\tp \ub)\nn\\
& \quad =
\cM_{1}^{\pm}+\tnp_{s}\cM_{2}^{\pm}+\tnp_{t}\cM_{3}^{\pm}+\tnp_{u}\cM_{4}^{\pm}\nn\\
&\qquad +i\sigma_{\mu\nu}\tp_{s}{}^{\mu}\tp_{t}{}^{\nu}\cM_{5}^{\pm}+i\sigma_{\mu\nu}\tp_{s}{}^{\mu}\tp_{u}{}^{\nu}\cM_{6}^{\pm}\nn\\
&\qquad +i\sigma_{\mu\nu}\tp_{t}{}^{\mu}\tp_{u}{}^{\nu}\cM_{7}^{\pm}\nn\\
&\qquad +i\gamma_{\mu}\gamma_{5}\varepsilon^{\mu\nu\rho\sigma}\tp_{s\nu}\tp_{t\rho}\tp_{u\sigma}\cM_{8}^{\pm}\,.
\end{align}
We use the convention $\varepsilon_{0123} = 1$.
Here the invariant amplitudes $\cM_{j}^{\pm}$ ($j=1, \dots , 8$)
can only depend on $\tilde{s}$, $\tilde{t}$ and the invariant squared masses,
\begin{align}\label{8}
&\cM_{j}^{\pm}=\cM_{j}^{\pm}(\tilde{s},\,\tilde{t},\,m_{1}\2,\, m_{2}\2,\, m\ua\2,\,m\ub\2)\,, \nn\\
& m_{a}^{2}=\tp_{a}^{2}\,,\quad
m_{b}^{2}=\tp_{b}^{2}\,,\quad
m_{1}^{2}=\tp_{1}^{2}\,,\quad
m_{2}^{2}=\tp_{2}^{2}\,.
\end{align}
%\newpage
%\noindent
Specializing \er{7} for the on-shell case we get back \er{6} with $\tilde{s}\to s$, $\ \tilde{t} \to t$, 
$\ m\ua\2 = m\uu\2 =m\upp\2$, $\ m\ub\2 =m\ud\2 = m\up\2$, 
and 
\begin{align}
\label{9}
A^{(\text{on})\pm}(s,t)=&\;\cM_{1}^{(\text{on})\pm}
+m_{p}\cM_{2}^{(\text{on})\pm}
-m_{p}\cM_{4}^{(\text{on})\pm}\nn\\
&+(-s+m_{p}\2 +m_{\pi}\2)\cM_{5}^{(\text{on})\pm}\nn\\
&+(s+t-m_{p}\2 -m_{\pi}\2)\cM_{7}^{(\text{on})\pm}\nn \\
&-m_{p}(2s+t-2m_{p}\2 -2m_{\pi}\2)\cM_{8}^{(\text{on})\pm}
\,,\\
\label{10}
B^{(\text{on})\pm}(s,t)=&\;\cM_{2}^{(\text{on})\pm}+\cM_{4}^{(\text{on})\pm}\nn\\
&+2m_{p}\cM_{5}^{(\text{on})\pm}-2m_{p}\cM_{7}^{(\text{on})\pm}\nn\\
&+(4m_{p}^{2}-t)\cM_{8}^{(\text{on})\pm}\,.
\end{align}
On shell the amplitudes $\cM_{3}^{\pm}$ and $\cM_{6}^{\pm}$ are zero from $C$ and $T$ invariance.

\begin{figure}[!ht]
(a)\includegraphics[width=.3\textwidth]{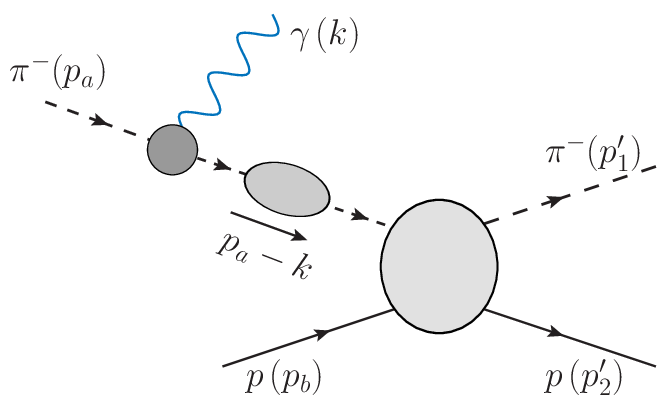}\\
(b)\includegraphics[width=.3\textwidth]{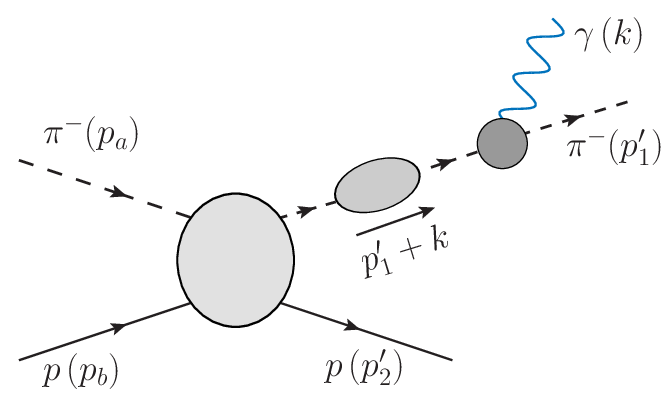}\\
(c)\quad \includegraphics[width=0.28\textwidth]{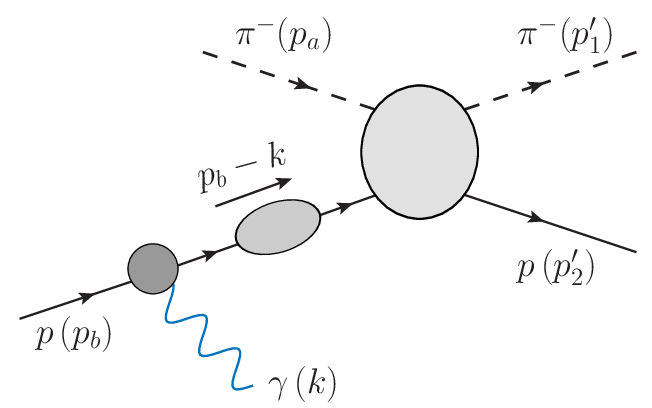}\\
(d)\includegraphics[width=.3\textwidth]{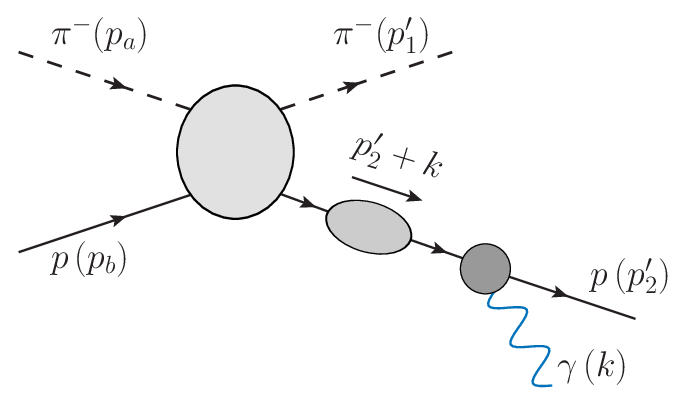}\\
(e)\qquad \includegraphics[width=.26\textwidth]{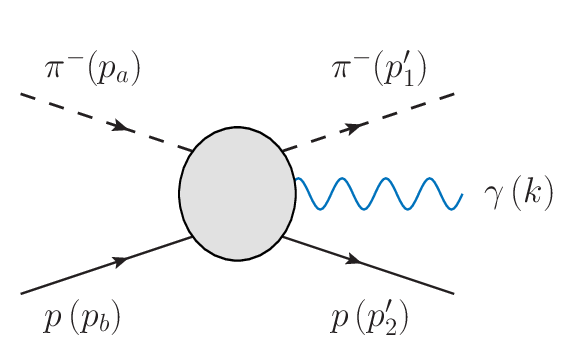}
\caption{Diagrams for the $\pi^{-} p \to \pi^{-} p \gamma$ reaction. Photon emission from external particles is shown in (a)--(d), the structure term (e) is non singular for $k \to 0$. The blobs in (a)--(d) represent the full propagators and vertices and the off-shell $\pi^{-} p \to \pi^{-} p$ amplitude.}
\label{fig:100}
\end{figure}

Next we consider the reactions \er{2}.
We have five diagrams for 
$\pi^{-} p\to \pi^{-} p\gamma$ 
as shown in Fig.~\ref{fig:100}. 
For $\pi^{+} p\to \pi^{+} p\gamma$ the diagrams are analogous.

Let $\cM\hpm\ul$ be the amplitude without spinors for \er{2}. 
We define a matrix amplitude $\cN\hpm\ul$ by
%\label{3.16} not same
\begin{align}
\label{16}
&\mathcal{N}\hpm\ul(p'\uu, p'\ud, k, p\ua ,p\ub)\nn\\
&\quad =(\slash{p}'\ud +m_{p})\cM\hpm\ul (p'\uu, p'\ud, k, p\ua ,p\ub)(\slash{p}\ub +m_{p})\,.
\end{align}
The ${\mathcal T}$-matrix element for \er{2} is then
\bal{17}
&\braket{\ppm (p'\uu), \, p(p'\ud , \lambda'\ud) ,\, \gamma (k,\varepsilon)|{\mathcal T}|\ppm (p\ua),\, p(p\ub , \lambda\ub )}\nn\\
&\quad =(\varepsilon^{\lambda})^{*} \frac{1}{(2m\up)\2}
\bar{u}(p'\ud , \lambda'\ud)
\cN\hpm\ul(p'\uu , p'\ud  , k , p\ua ,  p\ub) u(p\ub , \lambda\ub)\,.
\end{align}
The advantage of working with $\cN\hpm\ul$ 
instead of $\cM\hpm\ul$, sandwiched between spinors, 
is that we do not have to specify 
any particular spin basis for the protons. 

For real photon emission we have $k^{2} = 0$ in \er{17} and this is what we consider here. In \cite{Lebiedowicz:2023mlz} we treat the amplitude $\cN\hpm\ul$ \er{16} also for virtual photons, that is, for $k^{2} \neq 0$. 
The discussion of the kinematics of the reactions
(\ref{1}) and (\ref{2}) is analogous to the one for
(\ref{100}) and (\ref{101}) in Sec.~\ref{sec:2}.
We work again in the c.m. system where we have for (\ref{1})
\begin{eqnarray}
&& p_{a}^{0} = p_{1}^{0} = \frac{1}{2 \sqrt{s}}(s + m_{\pi}^{2} - m_{p}^{2})\,, \nn \\
&& p_{b}^{0} = p_{2}^{0} = \frac{1}{2 \sqrt{s}}(s - m_{\pi}^{2} + m_{p}^{2})\,, \nn \\
&& |\bm{p_{a}}| = |\bm{p_{b}}| = |\bm{p_{1}}| = |\bm{p_{2}}|
= \sqrt{(p_{a}^{0})^{2} - m_{\pi}^{2}}
= \sqrt{(p_{b}^{0})^{2} - m_{p}^{2}}\,.\nn\\
\label{7.100}
\end{eqnarray}
Given the initial state the phase space of the final state
for (\ref{1}) is parametrized by
$\bm{\hat{p}\uu}=\bm{p\uu}/|\bm{p\uu}|$
and for (\ref{2}) by
$(k, \bm{\hat{p}\uu'})$, where
$\bm{\hat{p}\uu'}=\bm{p\uu'}/|\bm{p\uu'}|$.
For small $\omega$ ($\omega \ll |\bm{p_{1}}|$)
$\bm{\hat{p}\uu'}$ can vary over the whole unit sphere.

We consider again a neighbourhood of a phase-space point
$(k=0,\bm{\hat{p}\uu})$ and set there
$\bm{\hat{p}\uu'} = \bm{\hat{p}\uu} - \bm{l_{1 \perp}}/|\bm{p_{1}}|$
with $|\bm{l_{1 \perp}}| = \mathcal{O}(\omega)$; see (\ref{4.1}).
This neighbourhood is then parametrized by $(k, \bm{l_{1 \perp}})$;
see Fig.~\ref{fig:5}.
For the momenta of the reaction (\ref{2}) 
at this phase-space point we set again 
$p'\uu=p\uu-l\uu$ and $p'\ud=p\ud-l\ud$,
where $l_{1,2}$ are determined up to order $\omega$
with the same result as in (\ref{15})
but inserting for $s$, $p_{1,2}$, $|\bm{p\uu}|$,
and $\bm{\hat{p}\uu} = \bm{p\uu}/|\bm{p\uu}|$,
the appropriate values for $\pi p$ scattering;
see (\ref{3})--(\ref{5}) and (\ref{7.100}).

Our aim is to derive the expansion of 
$\cN\hpm\ul (p\uu - l\uu, p\ud -l\ud, k, p\ua, p\ub)$
from~\er{16} for $\omega\to 0$ and to give the terms of order $\omega^{-1}$ and $\omega^{0}$ explicitly. 
Note that $l\uu$, $l\ud$, 
and $k$ are all of order~$\omega$; see~\er{15}. 
Thus, we have to expand $\cN\hpm\ul$ with respect to \textit{all} these momenta. Setting $l\uu = l\ud =0$ and expanding then only in $k$ makes no sense since this violates energy-momentum conservation and leads outside the physical region of the amplitude.
In the following we shall give the analog of the Laurent
expansion for $\pi^{-} \pi^{0} \to \pi^{-} \pi^{0} \gamma$
as discussed in Sec.~\ref{sec:4}.

In Figs.~\ref{fig:100}(a)--(d), the combinations of propagator times photon vertex occur for pion and proton. Using the generalized Ward identities 
we find for the pion (\ref{4.9}), 
see Fig.~\ref{fig:100}(a),
and for the proton in Fig.~\ref{fig:100}(c)
%\bal{B81}not same, but similar
\bal{19}
&S_{F}(p\ub -k)\, \widehat{\Gamma}^{(\gamma pp)\mu}(p\ub -k, p\ub )(\bp\ub +m\up)\nn\\
&\quad 
=\frac{\bp\ub +m\up- \! \bk}{-2p\ub \cdot k+k\2+i\varepsilon}\bigg{[}\gamma^{\mu}-\frac{i}{2m\up}\sigma^{\mu\nu}k_{\nu}F\ud(0)\bigg{]}\nn\\
&\qquad \times ( \bp\ub +m\up)+{\mathcal O}(\omega)\,.
\end{align}
Here $F\ud (0) = \mu\up / \mu_{N} - 1$ 
with $\mu\up$ the magnetic moment of the proton 
and $\mu_{N}$ the nuclear magneton.
Expressions similar to \er{4.9} and \er{19} apply for the vertex times propagator terms in Figs.~\ref{fig:100}(b) and \ref{fig:100}(d), respectively. 

We see from \er{4.9} and \er{19} that for the determination of the amplitudes $\cN\hpm\ul$ to the orders $\omega^{-1}$ and $\omega^{0}$ we have to know the off-shell amplitudes for $\ppm p \to \ppm p $ 
in Figs.~\ref{fig:100}(a)--(d) 
to the orders $\omega^{0}$ and $\omega^{1}$. 
Thus, we have to use \er{7} and make this expansion for the terms 
\mbox{$\tnp_{s}$},~\mbox{$\tnp_{t}$}, etc. as well as for the amplitudes 
$\cM\uu\hpm, \dots, \cM\hpm_{8}$. 
This expansion is \textit{different} 
for the terms corresponding to 
the diagrams of Fig.~\ref{fig:100}(a)--(d).
Finally, the structure term corresponding 
to Fig.~\ref{fig:100}(e) 
can be determined from the gauge-invariance constraint 
\bal{20}
k^{\lambda}\cN\ul 
= k^{\lambda}
\left(\cN\ul^{(a)}+\cN\ul^{(b)}+\cN\ul^{(c)}+\cN\ul^{(d)}+\cN\ul^{(e)} \right)=0\,.
\end{align}

After a long and rather complicated calculation we find that the result for the amplitudes $\cN\hpm\ul$ \er{16} 
for $\ppm p \to \ppm p \gamma$ to the orders $\omega^{-1}$ 
and $\omega^{0}$ can be expressed completely in terms of the on-shell amplitudes 
$A^{(\text{on})\pm}(s,t)$ and $B^{(\text{on})\pm}(s,t)$ for
$\ppm p \to \ppm p$ and their partial derivatives 
with respect to $s$ and $t$, 
%\label{3.41}
\begin{align}
\label{21}
A,_{s}^{\!\!(\text{on})\pm}(s,t)&=
\frac{\partial}{\partial s}A^{(\text{on})\pm}(s,t)\,,\nn\\
A,_{t}^{\!\!(\text{on})\pm}(s,t)&=
\frac{\partial}{\partial t}A^{(\text{on})\pm}(s,t)\,,\nn\\
B,_{s}^{\!\!(\text{on})\pm}(s,t)&=
\frac{\partial}{\partial s}B^{(\text{on})\pm}(s,t)\,,\nn\\
B,_{t}^{\!\!(\text{on})\pm}(s,t)&=
\frac{\partial}{\partial t}B^{(\text{on})\pm}(s,t)\,.
\end{align}
We find for real photon emission
\bel{22}
\cN\ul\hpm = \frac{1}{\omega}\widehat{\cN}\ul^{(0)\pm} +
\widehat{\cN}\ul^{(1)\pm} + {\mathcal O}(\omega)\,,
\ee
where with $j= 0, 1$
\bel{23}
\widehat{\cN}\ul^{(j)\pm} = \widehat{\cN}\ul^{(a+b+e1)(j)\pm} +\widehat{\cN}\ul^{(c+d+e2)(j)\pm}\,, 
\ee
\begin{widetext}
\bal{24}
\widehat{\cN}\ul^{(a+b+e1)(0)\pm}=&\pm e(\bp\ud+m\up)\bigg{[}A^{(\text{on})\pm}+\frac{1}{2}(\bp\ua +\bp\uu )B^{(\text{on})\pm}\Big {]}(\bp\ub +m\up)\;
\omega\bigg{[}-\frac{p_{a\lambda}}{p\ua\cdot k}+\frac{p_{1\lambda}}{p\uu\cdot k}\bigg{]}\,, \\
\label{25} 
\widehat{\cN}\ul^{(a+b+e1)(1)\pm}=
&\pm e(\bp\ud+m\up)\bigg{[}A^{(\text{on})\pm}+\frac{1}{2}(\bp\ua +\bp\uu )B^{(\text{on})\pm}\Big {]}(\bp\ub +m\up)\frac{1}{(p\uu\cdot k)\2}\bigg{[}p_{1\lambda}(l\uu\cdot k)-l_{1\lambda}(p\uu\cdot k)\bigg{]}\nn\\
&\pm e(-\!\not{l}\ud)\bigg{[}A^{(\text{on})\pm}+\frac{1}{2}(\bp\ua +\bp\uu )B^{(\text{on})\pm}\Big {]}(\bp\ub +m\up)
\bigg{[}-\frac{p_{a\lambda}}{p\ua \cdot k}+\frac{p_{1\lambda}}{p\uu \cdot k}\bigg{]}\nn\\
&\pm e(\bp\ud+m\up) \frac{1}{2}(-\!\not{l}\uu)B^{(\text{on})\pm} (\bp\ub +m\up)
\bigg{[}-\frac{p_{a\lambda}}{p\ua \cdot k}+\frac{p_{1\lambda}}{p\uu \cdot k}\bigg{]}\nn\\
&\pm e (\bp\ud +m\up)\bigg{\lbrace}\bigg{[}A,_{s}^{\!\!(\text{on})\pm}
+\frac{1}{2}(\bp\ua+\bp\uu)
B,_{s}^{\!\!(\text{on})\pm}
\bigg{]}
\bigg{[} 2(p\us \cdot k) \frac{p_{a\lambda}}{p\ua\cdot k}-2p_{s\lambda}\bigg{]}\nn\\
&+\bigg{[}A,_{t}^{\!\!(\text{on})\pm}+\frac{1}{2}(\bp\ua+\bp\uu)B,_{t}^{\!\!(\text{on})\pm}\bigg{]}
2(p\ut \cdot l\ud)
\bigg{[}
\frac{p_{a\lambda}}{p\ua \cdot k}-\frac{p_{1\lambda}}{p\uu \cdot k}\bigg{]}\nn\\
&+B^{(\text{on})\pm}\bigg{[}\frac{1}{2}\bk \bigg{(}\frac{p_{a\lambda}}{p\ua \cdot k}+\frac{p_{1\lambda}}{p\uu \cdot k}\bigg{)}-\gamma_{\lambda}\bigg{]}\bigg{\rbrace}(\bp\ub+m\up)
\,,\\
%\end{align}
\label{26}
\widehat{\cN}\ul^{(c+d+e2)(0)\pm}=&\;
e(\bp\ud+m\up)
\bigg{[}A^{(\text{on})\pm}+\frac{1}{2}(\bp\ua +\bp\uu)
       B^{(\text{on})\pm}\Big {]}(\bp\ub +m\up)\;
\omega\bigg{[}
-\frac{p_{b\lambda}}{p\ub \cdot k}+\frac{p_{2\lambda}}{p\ud \cdot k}\bigg{]}\,, 
\\
\label{27}
\widehat{\cN}\ul^{(c+d+e2)(1)\pm}=&\;
e(\bp\ud+m\up)
\bigg{[}A^{(\text{on})\pm}+\frac{1}{2}(\bp\ua +\bp\uu )B^{(\text{on})\pm}\Big {]}(\bp\ub +m\up)
\frac{1}{(p\ud\cdot k)\2}\bigg{[}p_{2\lambda} (l\ud \cdot k )-l_{2\lambda}(p\ud \cdot k)\bigg{]}\nn\\
&+e(-\!\not{l}\ud)
\bigg{[}A^{(\text{on})\pm}+\frac{1}{2}(\bp\ua +\bp\uu )B^{(\text{on})\pm}\Big {]}(\bp\ub +m\up)
\bigg{[}
-\frac{p_{b\lambda}}{p\ub \cdot k}+\frac{p_{2\lambda}}{p\ud \cdot k}\bigg{]}
\nn\\
&+e(\bp\ud +m\up)
\frac{1}{2}(-\!\not{l}\uu) B^{(\text{on})\pm}(\bp\ub +m\up)\bigg{[}
-\frac{p_{b\lambda}}{p\ub \cdot k}+\frac{p_{2\lambda}}{p\ud \cdot k}\bigg{]}
\nn\\
&+e(\bp\ud +m\up)
\bigg{[}A,_{s}^{\!\!(\text{on})\pm}+\frac{1}{2}(\bp\ua+\bp\uu)B,_{s}^{\!\!(\text{on})\pm}\bigg{]}(\bp\ub +m\up)
\bigg{[}2(p\us\cdot k)\frac{p_{b\lambda}}{p\ub \cdot k}-2p_{s \lambda}\bigg{]}
\nn\\
&+e(\bp\ud +m\up)\bigg{[}A,_{t}^{\!\!(\text{on})\pm}+\frac{1}{2}(\bp\ua+\bp\uu)B,_{t}^{\!\!(\text{on})\pm}\bigg{]}(\bp\ub +m\up)
2(p\ut \cdot l\uu)\bigg{[}-\frac{p_{b\lambda}}{p\ub \cdot k}+\frac{p_{2\lambda}}{p\ud \cdot k}\bigg{]}\nn \allowdisplaybreaks\\
&+e(\bp\ud +m\up)
\bigg{[}A^{(\text{on})\pm}+\frac{1}{2}(\bp\ua+\bp\uu)B^{(\text{on})\pm}\bigg{]}
(k\ul -\bk \gamma\ul)(\bp\ub +m\up)\frac{1}{(-2p\ub\cdot k)}\nn \\
&-e\frac{1}{(2p\ud\cdot k)}(\bp\ud +m\up)(k\ul -\gamma\ul \bk)
\bigg{[}A^{(\text{on})\pm}+\frac{1}{2}(\bp\ua+\bp\uu)B^{(\text{on})\pm}\bigg{]}(\bp\ub +m\up)\nn\\
&+e(\bp\ud + m\up)\bigg{[}A^{(\text{on})\pm}+\frac{1}{2}(\bp\ua+\bp\uu)B^{(\text{on})\pm}\bigg{]}
\bigg{[}
m\up(k\ul-\bk\gamma\ul)+\bigg{(}p_{b\lambda}\bk -(p\ub\cdot k )\gamma\ul\bigg{)}\bigg{]}(\bp\ub+m\up) \nn\\
&\quad \times 
\frac{F\ud(0)}{m\up}\frac{1}{(-2p\ub\cdot k)}
%\nn\\ &
-e\frac{F\ud(0)}{m\up}\frac{1}{(2p\ud\cdot k)}(\bp\ud + m\up)\bigg{[} m\up (k\ul -\gamma\ul \bk)+\bigg{(}p_{2\lambda}\bk -(
p\ud \cdot k)\gamma\ul \bigg{)}\bigg{]}\nn\\
&\quad \times \bigg{[}A^ {(\text{on})\pm}+\frac{1}{2}(\bp\ua+\bp\uu)B^ {(\text{on})\pm}\bigg{]}(\bp\ub +m\up)\,.
\end{align}
\end{widetext}

When our calculations for the amplitudes for 
$\pi^{\pm} p \to \pi^{\pm} p \gamma$ in the soft-photon limit
were finished we learned about 
Refs.~\cite{Liou:1978jx, Liou:1987ug, Liou:1977qv, Liou:1978sz, Liou:1982fm}.
There, as in our work, it is emphasized that one has to make
a consistent expansion of all terms
in the amplitude for $\omega \to 0$.
But our results rely on more general premisses.
We consider, for instance, the most general off-shell
$\pi p \to \pi p$ amplitudes
which contain eight invariant amplitudes;
see (\ref{7}).
In \cite{Liou:1978sz} only two invariant
off-shell amplitudes are considered.
A detailed comparison of our methods and results
with those of
\cite{Liou:1977qv, Liou:1978sz, Liou:1982fm}
shall be given in an update of \cite{Lebiedowicz:2023mlz}.

For $\pi^{0} p \to \pi^{0} p \gamma$ 
we are left with the diagrams (c), (d), and (e) 
of Fig.~\ref{fig:100} with $\pi^{-} \to \pi^{0}$. 
Therefore, in our method, the amplitude
$\cN^{0}\ul$ for $\pi^{0} p \to \pi^{0} p \gamma$ 
to the orders $\omega^{-1}$ and $\omega^{0}$ 
for real photon emission can be expressed as
\bal{7.21}
&\cN\ul^{0} = \frac{1}{\omega}\widehat{\cN}\ul^{(0)0} +
\widehat{\cN}\ul^{(1)0} + {\mathcal O}(\omega)\,,
\intertext{with}
&\widehat{\cN}\ul^{(j)0} = \widehat{\cN}\ul^{(c+d+e2)(j)0}\,,
\quad j = 0,1\,,
\label{7.22}
\end{align}
using expressions analogous to (\ref{26}) and (\ref{27})
replacing, of course, the $\pi^{\pm} p$
by the $\pi^{0} p$ amplitudes.
Again we emphasize that Low's formula, 
(3.16) of \cite{Low:1958sn}, for the radiative amplitude
for $\pi^{0} p \to \pi^{0} p \gamma$
gives an approximate result valid at the fixed
phase-space point \mbox{($p_{1}'$, $p_{2}'$, $k$)},
while our result (\ref{7.22}) corresponds
to the Laurent expansion in $\omega$ around
the phase-space point \mbox{($p_{1}$, $p_{2}$, $k=0$)}.

%--------------------------------------------------
\section{Conclusions}
\label{sec:8}
%--------------------------------------------------

In this article we have first discussed $\pi^{-}\pi^{0}$ elastic scattering 
and soft-photon radiation in this reaction.
We have shown that the soft-photon theorems of Low \cite{Low:1958sn}
and Weinberg \cite{Weinberg:1964ew,Weinberg:1965nx} have a different meaning.
In \cite{Low:1958sn} an approximate expression for the radiative amplitude
at a \textit{given} phase-space point 
\mbox{($p_{1}'$, $p_{2}'$, $k$)} is presented.
In \cite{Weinberg:1965nx} the pole term of the radiative amplitude
in a Laurent expansion in the photon energy $\omega$ around
the phase-space point \mbox{($p_{1}$, $p_{2}$, $k=0$)} is given,
where $p_{1}$, $p_{2}$ are the momenta of the non-radiative reaction.
We have recalled and discussed our calculation of \cite{Lebiedowicz:2021byo} 
where we presented the $\omega^{0}$ term in the above Laurent expansion.
We have discussed in detail that the next-to-leading terms
presented in \cite{Low:1958sn} and \cite{Lebiedowicz:2021byo},
respectively, must be different,
since their meaning is different.
We have then constructed the Laurent expansion of Low's formula (\ref{5.4})
around the phase-space point \mbox{($p_{1}$, $p_{2}$, $k=0$)}
and found complete agreement with our results of \cite{Lebiedowicz:2021byo}.
We emphasized that Low's formula (\ref{5.4}) for 
the radiative amplitude ${\cal M}_{\lambda}$
is \textit{only} valid for the \textit{one physical value}
of \mbox{$k = p_{a} + p_{b} - p_{1}' - p_{2}'$}.
In contrast, in (\ref{4.13}) we give a Laurent expansion of ${\cal M}_{\lambda}$
around the phase-space point ($p_{1}$, $p_{2}$, $k=0$) where,
of course, we can vary the expansion parameters $(k, \bm{l_{1 \perp}})$
which are defined at the beginning of Sec.~\ref{sec:4}.

In \cite{Lebiedowicz:2021byo} we wrongly interpreted Low's formula 
as an expansion of ${\cal M}_{\lambda}$ 
around ($p_{1}$, $p_{2}$, $k=0$).
With this premiss we came to the conclusion that Low's formula
violates energy-momentum conservation. 
We have now discussed in detail
that the above premiss does not hold.
Thus, also our above conclusion does not hold.
Therefore, the part of Sec.~III of \cite{Lebiedowicz:2021byo} after Eq.~(3.29) is obsolete
since it was based on a wrong premiss.
This was already explained in \cite{Nachtmann_talk}
and will be subject of a forthcoming erratum
to \cite{Lebiedowicz:2021byo}.
We were confused by the fact that in the literature frequently
Weinberg's formula (\ref{4.14}) is referred to as Low's formula.
But, as we hope to have demonstrated in the present paper,
these two formulas have a \textit{different} meaning.
This was the subject of our Secs.~\ref{sec:4} and \ref{sec:5}.

In Sec.~\ref{sec:6} we discussed the expansions of different cross sections
of $\pi^{-}(p_{a}) + \pi^{0}(p_{b}) 
\to \pi^{-}(p_{1}') + \pi^{0}(p_{2}') + \gamma(k)$ for $\omega \to 0$.
We found that for calculating these expansions for 
\mbox{$\omega \dv\sigma/(\dv \omega \, \dv \Omega_{\hat{k}} \, \dv \Omega_{\hat{p}_{1}'})$} and 
\mbox{$\omega \dv\sigma/(\dv \omega \, \dv \Omega_{\hat{k}} \, \dv \Omega_{\hat{p}_{2}'})$} 
it was necessary to have at our disposal the expansion of
the radiative amplitude in \textit{all} directions 
$(k, \bm{\hat{p}_{1}'})$ of the phase space around
the non-radiative point $(k = 0, \bm{\hat{p}_{1}})$.
We calculated then the expansion in $\omega$ for $\omega \to 0$
of the cross section \mbox{$\omega \dv\sigma/\dv \omega$}
from the above two differential cross sections.
As it must be, we found the same result using these two ways.

Finally we have discussed in Sec.~\ref{sec:7} the soft-photon expansion
of the amplitudes for the reactions $\pi^{\pm} p \to \pi^{\pm} p \gamma$.
We have presented a strict theorem of QCD. 
Given the amplitudes for $\ppm p \to \ppm p$ scattering 
the amplitudes for soft-photon production, 
$\ppm p \to \ppm p \gamma$, 
have been calculated exactly 
to the orders $\omega^{-1}$ and $\omega^{0}$. 
These two orders of the expansion of the 
$\ppm p \to \ppm p \gamma$ amplitudes 
are completely determined by the $\ppm p \to \ppm p$ on-shell amplitudes. 
For real photon emission the result is given in \er{22}--\er{27}. 
The results for soft virtual photon emission and
consequences for cross sections are given 
in \cite{Lebiedowicz:2023mlz}.

The derivation of our results, especially for the $\ppm p \to \ppm p \gamma$
reaction, involved lengthy calculations.
Thus, we found it convenient to check our general results in a model
which satisfies the QFT constraints for these reactions
as listed at the end of Sec.~\ref{sec:1}.
Such a model is the tensor-Pomeron model of \cite{Ewerz:2013kda}
but improved for reactions involving photons as shown
in Sec.~IV~A of \cite{Lebiedowicz:2021byo} 
for $\pi^{-} \pi^{0} \to \pi^{-} \pi^{0} \gamma$
and in \cite{Lebiedowicz:2022nnn,Lebiedowicz:2023mhe} 
for $pp \to pp \gamma$.
Applying this model as done in \cite{Lebiedowicz:2021byo} to 
$\pi^{-} \pi^{0} \to \pi^{-} \pi^{0} \gamma$ and expanding
the resulting expressions [Eqs.~(4.19), (4.22)--(4.24) from \cite{Lebiedowicz:2021byo}]
in $\omega$ for $\omega \to 0$ we get indeed 
the result expected from
(3.27), (3.28), and (A1),
of \cite{Lebiedowicz:2021byo} 
which, for real photon emission,
we reproduce in 
(\ref{4.11})--(\ref{4.13}) of the present paper.
We also checked that this improved tensor-Pomeron model gives
amplitudes for $\pi^{\pm} p \to \pi^{\pm} p \gamma$
where the expansion in $\omega$ agrees with the general results
(\ref{21})--(\ref{27}) in our present paper.
The details for this will be discussed elsewhere.

To summarize: we hope to have clarified the meaning of the soft-photon
expansions in the versions of F.E.~Low~\cite{Low:1958sn}
and S.~Weinberg~\cite{Weinberg:1964ew,Weinberg:1965nx}.
These expansions must not be confounded, 
as they have a \textit{different} meaning.
We have shown how these two expansions are related.
We have discussed the Laurent expansions in the photon energy $\omega$
for $\omega \to 0$ for the reactions
$\pi^{-} \pi^{0} \to \pi^{-} \pi^{0} \gamma$ and
$\pi^{\pm} p \to \pi^{\pm} p \gamma$ to the orders
$\omega^{-1}$ (the pole term of \cite{Weinberg:1965nx}) and $\omega^{0}$.
Our results are strict consequences of QCD.
We hope that they will be helpful also for experimentalists
who are embarked to check soft-photon theorems.
For $\pi^{\pm} p \to \pi^{\pm} p \gamma$ this could be done perhaps at COMPASS
and 
for $\pi^{\pm} p \to \pi^{\pm} p \gamma^{*} (\to e^{+}e^{-})$
in HADES at GSI \cite{Rathod:2022pof}.
For ALICE~3 \cite{ALICE:2022wwr} 
we would need the corresponding theoretical calculations
for $pp \to pp \gamma$.
Our methods are suited to calculate 
in a rigorous way the expansion of the amplitude 
to the orders $\omega^{-1}$ and $\omega^{0}$.
These calculations certainly will not be easy.

After our paper was finished there appeared on the arXiv
the paper \cite{Fadin:2024tar} 
where our work of \cite{Lebiedowicz:2021byo} was criticized.
Some of this criticism is acceptable, but we have already
discussed our misinterpretation
of the soft-photon theorem of \cite{Low:1958sn} in
Ref.~\cite{Nachtmann_talk}.
In our present paper we again discuss extensively this
misinterpretation. But, in this connection,
a main topic of our paper is to show that the soft-photon theorems
of \cite{Weinberg:1964ew,Weinberg:1965nx}
(to which there is no reference in \cite{Fadin:2024tar})
and of \cite{Low:1958sn} are different
and should not be confounded.
Of course, they are related, as we have discussed in Sec.~\ref{sec:5}.

\begin{acknowledgments}
We are very grateful to an anonymous referee of a first version
of this paper and a related paper by us for pointing out
our misinterpretation of Ref.~\cite{Low:1958sn}.
We thank J.~Stachel, P.~Braun-Munzinger, and C.~Ewerz 
for discussions and useful suggestions.
This work was partially supported by
the Polish National Science Centre under Grant
No. 2018/31/B/ST2/03537.
\end{acknowledgments}

%--------------------
\appendix

\renewcommand\theequation{\thesection\arabic{equation}} 
%--------------------
\section{The expansion of $1/p_{1}' \cdot k$}
\label{sec:appendixA}

In this Appendix we discuss the region of validity
of the expansion (\ref{4.12}).
For this we have to consider 
\mbox{$|l_{1} \cdot k| / |p_{1} \cdot k|$}.
Working in the c.m. system we have
[see (\ref{2.3}), (\ref{15}) and (\ref{4.6})]
\begin{eqnarray}
p_{1} \cdot k &=& p_{1}^{0} \omega - \bm{p_{1}} \cdot \bm{k} =
\omega  ( p_{1}^{0} - |\bm{p_{1}}| \bm{\hat{p}_{1}} \cdot \bm{\hat{k}} )
 \nn \\
&=& 
\omega \bigg{[} \frac{m_{\pi}^{2}}{p_{1}^{0} + |\bm{p_{1}}|}
+ |\bm{p_{1}}| (1 - \bm{\hat{p}_{1}} \cdot \bm{\hat{k}}) \bigg{]}
\,, \nn\\
l_{1} \cdot k &=& 
l_{1}^{0} \omega - \bm{l_{1}} \cdot \bm{k} =
\omega  ( l_{1}^{0} - \bm{l_{1}} \cdot \bm{\hat{k}} ) 
\nn \\
&=& 
\omega \bigg{[}
\frac{p_{2} \cdot k}{2} 
\bigg{(}
\frac{1}{p_{1}^{0}} - \frac{1}{|\bm{p_{1}}|}
\bm{\hat{p}_{1}} \cdot \bm{\hat{k}}
\bigg{)}
- \bm{l_{1 \perp}} \cdot \bm{\hat{k}}
\bigg{]}
\nn \\
&=& 
\omega \bigg{[}
\frac{p_{2} \cdot k}{2} 
\frac{|\bm{p_{1}}| - p_{1}^{0}}{p_{1}^{0} |\bm{p_{1}}|} 
+ \frac{p_{2} \cdot k}{2 |\bm{p_{1}}|} 
(1 - \bm{\hat{p}_{1}} \cdot \bm{\hat{k}})
- \bm{l_{1 \perp}} \cdot \bm{\hat{k}}
\bigg{]}
\nn \\
&=& 
\omega^{2} \bigg{[}
- \frac{m_{\pi}^{2} (p_{1}^{0} + \bm{\hat{p}_{1}} \cdot \bm{\hat{k}} \,|\bm{p_{1}}|)}{2 (p_{1}^{0} + |\bm{p_{1}}|)\, p_{1}^{0} |\bm{p_{1}}|} \nn \\
&&+ \frac{p_{1}^{0} 
+ \bm{\hat{p}_{1}} \cdot \bm{\hat{k}} \,|\bm{p_{1}}|}{2 |\bm{p_{1}}|}
(1 - \bm{\hat{p}_{1}} \cdot \bm{\hat{k}}) 
- \bm{\tilde{l}_{1 \perp}} \cdot \bm{\hat{k}}
\bigg{]} \,.\quad
\label{A1}
\end{eqnarray}

From (\ref{A1}) we get
\begin{eqnarray}
\frac{|l_{1} \cdot k|}{|p_{1} \cdot k|} & \leqslant & \frac{\omega}{|\bm{p_{1}}|}
\bigg{[} \frac{m_{\pi}^{2}}{p_{1}^{0} + |\bm{p_{1}}|}
(2 - \bm{\hat{p}_{1}} \cdot \bm{\hat{k}}) 
+ |\bm{p_{1}}| (1 - \bm{\hat{p}_{1}} \cdot \bm{\hat{k}}) \nn\\
&& + |\bm{p_{1}}| |\bm{\tilde{l}_{1 \perp}} \cdot \bm{\hat{k}}|
\bigg{]}
\bigg{[} \frac{m_{\pi}^{2}}{p_{1}^{0} + |\bm{p_{1}}|}
+ |\bm{p_{1}}|
(1 - \bm{\hat{p}_{1}} \cdot \bm{\hat{k}}) 
\bigg{]}^{-1}. \nn\\
\label{A2}
\end{eqnarray}

For $\bm{\tilde{l}_{1 \perp}} = 0$ we have, therefore,
\begin{eqnarray}
\frac{|l_{1} \cdot k|}{|p_{1} \cdot k|} =
{\cal O}\left( \frac{\omega}{|\bm{p_{1}}|} \right)
\label{A3}
\end{eqnarray}
and the expansion (\ref{4.12}) is valid for
$\omega \ll |\bm{p_{1}}|$ and all $\bm{\hat{p}_{1}} \cdot \bm{\hat{k}}$.

For $|\bm{\tilde{l}_{1 \perp}} \cdot \bm{\hat{k}}| \neq 0$ and
$1 - \bm{\hat{p}_{1}} \cdot \bm{\hat{k}} = {\cal O}(1)$ 
we still get (\ref{A3}).
But for $\bm{\hat{p}_{1}} \cdot \bm{\hat{k}} = 1$
we find from (\ref{A2}) only
\begin{eqnarray}
\frac{|l_{1} \cdot k|}{|p_{1} \cdot k|} \leqslant
\omega \bigg{[}
\frac{1}{|\bm{p_{1}}|}
+
\frac{p_{1}^{0} + |\bm{p_{1}}|}
{m_{\pi}^{2}}
|\bm{\tilde{l}_{1 \perp}} \cdot \bm{\hat{k}}|\bigg{]}\,.
\label{A4}
\end{eqnarray}

In this case we have to require for the expansion (\ref{4.12}) to be valid
\begin{eqnarray}
\omega \leqslant \frac{m_{\pi}^{2} |\bm{p_{1}}|}{ m_{\pi}^{2}+
|\bm{p_{1}}| (p_{1}^{0}+|\bm{p_{1}}|)\,
|\bm{\tilde{l}_{1 \perp}} \cdot \bm{\hat{k}}|}\,,
\label{A5}
\end{eqnarray}
\\
which is very small for momenta $|\bm{p_{1}}| \gg m_{\pi}$.

To conclude: for $p_{1}' = p_{1} - l_{1}$ the expansion
\begin{eqnarray}
\frac{1}{(p_{1} - l_{1}, k)} = \frac{1}{p_{1} \cdot k}
\bigg{[} 1 + \frac{l_{1} \cdot k}{p_{1} \cdot k} +
{\cal O}(\omega^{2}) \bigg{]}
\label{A6}
\end{eqnarray}
and thus (\ref{4.12}) is alright under the following conditions.
%-----------------------------------
\begin{itemize}
%-----------------------------------
\item[(i)] 
%-----------------------------------
For $\bm{\hat{p}\uu'} = \bm{\hat{p}\uu}$
in the c.m. system we have to require from (\ref{A3})
\begin{eqnarray}
\omega \ll |\bm{p_{1}}| \,.
\label{A7}
\end{eqnarray}
%-----------------------------------
\item[(ii)] 
%-----------------------------------
For $\bm{\hat{p}\uu'} 
= \bm{\hat{p}\uu} - \bm{l_{1  \perp}}/|\bm{p_{1}}|$ and 
$\bm{\tilde{l}_{1 \perp}} \cdot \bm{\hat{k}} = 0$
the requirement is again (\ref{A7}).
For $\bm{\tilde{l}_{1 \perp}} \cdot \bm{\hat{k}} \neq 0$
we get here from (\ref{A2}) the condition
\begin{eqnarray}
\omega & \ll & |\bm{p_{1}}|
\bigg{[} \frac{m_{\pi}^{2}}{p_{1}^{0} + |\bm{p_{1}}|}
+ |\bm{p_{1}}| (1 - \bm{\hat{p}_{1}} \cdot \bm{\hat{k}}) 
\bigg{]} \nn\\
&& \times
\bigg{[} \frac{m_{\pi}^{2}}{p_{1}^{0} + |\bm{p_{1}}|}
(2 - \bm{\hat{p}_{1}} \cdot \bm{\hat{k}}) 
+ |\bm{p_{1}}|
(1 - \bm{\hat{p}_{1}} \cdot \bm{\hat{k}}) \nn \\
&& \quad
+ |\bm{p_{1}}| |\bm{\tilde{l}_{1 \perp}} \cdot \bm{\hat{k}}|
\bigg{]}^{-1}.
\label{A8}
\end{eqnarray}
For $1 - \bm{\hat{p}_{1}} \cdot \bm{\hat{k}} \neq 0$ and of order 1
we get again (\ref{A7}).
But for $\bm{\hat{p}_{1}} \cdot \bm{\hat{k}} = 1$ (\ref{A8})
gives (\ref{A5})
which is very small for 
$|\bm{\tilde{l}_{1 \perp}} \cdot \bm{\hat{k}}| = {\cal O}(1)$ and
$|\bm{p_{1}}| \gg m_{\pi}$.

%-----------------------------------
\end{itemize}
%-----------------------------------

%--------------------
\section{Variable transformation}
\label{sec:appendixB}

We consider here the variable transformation (\ref{6.2b})
$\bm{\hat{p}_{1}'} \to \bm{\hat{p}_{1}''}$
for fixed $\bm{\hat{k}}$ and $\omega$:
\begin{align}
\hat{p}_{1i}' = \hat{p}_{1i}''
- \frac{\omega}{|\bm{p_{1}}|}
\left[
\hat{k}_{i} \hat{p}_{1j}'' \hat{p}_{1j}'' - 
\hat{p}_{1j}'' \hat{k}_{j}
\hat{p}_{1i}''
\right] \,.
\label{B1}
\end{align}
We have
\begin{eqnarray}
&&(\bm{\hat{p}_{1}'})^{2} = (\bm{\hat{p}_{1}''})^{2} +
{\cal O}(\omega^{2})\,,
\label{B2} \\
&&\frac{\partial \hat{p}_{1i}'}{\partial \hat{p}_{1e}''}
= \delta_{ie} + \frac{\omega}{|\bm{p_{1}}|}
\left[
\hat{p}_{1i}'' \hat{k}_{e} +
\bm{\hat{p}_{1}''} \cdot \bm{\hat{k}}  \delta_{ie}
-2 \hat{k}_{i} \hat{p}_{1e}''
\right], \qquad
\label{B3} \\
&& {\rm det} \left( \frac{\partial \hat{p}_{1i}'}{\partial \hat{p}_{1e}''} \right) = 1 + \frac{2 \omega}{|\bm{{p}_{1}}|} 
\bm{\hat{p}_{1}''} \cdot \bm{\hat{k}} +
{\cal O}(\omega^{2})\,.
\label{B4} 
\end{eqnarray}

\begin{widetext}

Now we consider an arbitrary function 
$f(\bm{\hat{p}_{1}'})$ setting
\begin{align}
\tilde{f}(\bm{\hat{p}_{1}''}) = f(\bm{\hat{p}_{1}'})\,.
\label{B5}
\end{align}
We are interested in the following integral and its variable transformed:
\begin{eqnarray}
\int 
\dv \Omega_{\hat{p}_{1}'}
J^{(1)}(s, \omega, \bm{\hat{p}\uu'}, \bm{\hat{k}}) \,
f(\bm{\hat{p}_{1}'})
&=&
2 \int \dv^{3} \hat{p}_{1}' \, 
\delta((\bm{\hat{p}_{1}'})^{2} - 1)
J^{(1)}(s, \omega, \bm{\hat{p}\uu'}, \bm{\hat{k}}) \,
f(\bm{\hat{p}_{1}'}) \nn \\
&=&
2 \int \dv^{3} \hat{p}_{1}'' \, 
{\rm det} \left( \frac{\partial \hat{p}_{1i}'}{\partial \hat{p}_{1e}''} \right)
\delta((\bm{\hat{p}_{1}''})^{2} - 1)
J^{(1)}(s, \omega, \bm{\hat{p}\uu'}, \bm{\hat{k}}) 
\tilde{f}(\bm{\hat{p}_{1}''}) 
+
{\cal O}(\omega^{2})
\nn \\
&=&
2 \int \dv^{3} \hat{p}_{1}'' \, 
\delta((\bm{\hat{p}_{1}''})^{2} - 1)
\left[ 1 + \frac{2 \omega}{|\bm{p_{1}}|} \bm{\hat{p}_{1}''} \cdot \bm{\hat{k}} \right]
J^{(1)}(s, \omega, \bm{\hat{p}\uu'}, \bm{\hat{k}}) 
\tilde{f}(\bm{\hat{p}_{1}''}) +
{\cal O}(\omega^{2})\,.
\label{B6}
\end{eqnarray}

From (\ref{6.5}) we find
\begin{eqnarray}
\left[ 1 + \frac{2 \omega}{|\bm{{p}_{1}}|} 
\bm{\hat{p}_{1}''} \cdot \bm{\hat{k}} \right]
J^{(1)}(s, \omega, \bm{\hat{p}\uu'}, \bm{\hat{k}})
&=&
\frac{1}{2} \sqrt{1 - \frac{4 m_{\pi}^{2}}{s}}
- \frac{\omega}{\sqrt{s}}
\bigg{(}
\frac{2 m_{\pi}^{2}}{\sqrt{s(s-4 m_{\pi}^{2})}} + \bm{\hat{p}\uu'} \cdot \bm{\hat{k}}
\bigg{)} + \frac{1}{2} \sqrt{1 - \frac{4 m_{\pi}^{2}}{s}}
\frac{2 \omega}{|\bm{{p}_{1}}|} \bm{\hat{p}_{1}''} \cdot \bm{\hat{k}} +
{\cal O}(\omega^{2}) \nn \\
&=&
\frac{1}{2} \sqrt{1 - \frac{4 m_{\pi}^{2}}{s}}
- \frac{\omega}{\sqrt{s}}
\frac{2 m_{\pi}^{2}}{\sqrt{s(s-4 m_{\pi}^{2})}} + 
\frac{\omega}{\sqrt{s}}
\bm{\hat{p}_{1}''} \cdot \bm{\hat{k}}
+ {\cal O}(\omega^{2}) \nn \\
&=&
J^{(1)}(s, \omega, -\bm{\hat{p}\uu''}, \bm{\hat{k}}) +
{\cal O}(\omega^{2}) \nn \\
&=&
J^{(1)}(s, \omega, \bm{\hat{p}\ud'}, \bm{\hat{k}}) +
{\cal O}(\omega^{2}) \,,
\label{B7} 
\end{eqnarray}
where we use $\bm{\hat{p}\ud'} = -\bm{\hat{p}\uu''}$;
see (\ref{6.2a1}).

Inserting (\ref{B7}) in (\ref{B6}) we obtain
\begin{eqnarray}
\int 
\dv \Omega_{\hat{p}_{1}'}
J^{(1)}(s, \omega, \bm{\hat{p}\uu'}, \bm{\hat{k}}) \,
f(\bm{\hat{p}_{1}'})=
\int 
\dv \Omega_{\hat{p}_{1}''}
J^{(1)}(s, \omega, -\bm{\hat{p}\uu''}, \bm{\hat{k}}) \,
\tilde{f}(\bm{\hat{p}_{1}''}) +
{\cal O}(\omega^{2})\,.
\label{B8}
\end{eqnarray}
Since the function $f(\bm{\hat{p}_{1}'})$ was arbitrary, 
we get the following transformation of the measures:
\begin{eqnarray}
\dv \Omega_{\hat{p}_{1}'}
J^{(1)}(s, \omega, \bm{\hat{p}\uu'}, \bm{\hat{k}})
&=&
\dv \Omega_{\hat{p}_{1}''}
J^{(1)}(s, \omega, -\bm{\hat{p}\uu''}, \bm{\hat{k}}) +
{\cal O}(\omega^{2}) \nn\\
&=&
\dv \Omega_{\hat{p}_{2}'}
J^{(1)}(s, \omega, \bm{\hat{p}\ud'}, \bm{\hat{k}}) +
{\cal O}(\omega^{2})\,.
\label{B9}
\end{eqnarray}

\end{widetext}

% Create the reference section using BibTeX:\clearpage
\bibliography{refs}

\end{document}